\documentclass[12pt]{article}
\usepackage{latexsym}
\usepackage{epsfig, amssymb, amscd, euscript, slashed}
\usepackage{amsmath}
\usepackage{array,calc,epsfig}
\usepackage[linktocpage]{hyperref}
\usepackage{pstricks}
\usepackage{pgf}
\usepackage{bbm}
\usepackage{fancybox}
\usepackage{tikz}
\usetikzlibrary{matrix,arrows,decorations.pathmorphing}

\usepackage{pgf}
\usepackage{tikz}
\usetikzlibrary{shapes,arrows}
\usepackage{pgf}
\usepgfmodule{shapes}
\usepgfmodule{plot}
\usetikzlibrary{decorations}
\usetikzlibrary{arrows}
\usetikzlibrary{snakes}

\tikzstyle{decision} = [diamond, draw,
    text width=4.5em, text badly centered, node distance=3cm, inner sep=0pt]
\tikzstyle{block} = [rectangle, draw,
    text width=5em, text centered, rounded corners, minimum height=4em]
\tikzstyle{line} = [draw, -latex']
\tikzstyle{cloud} = [draw, ellipse]

\def\ie{{\it i.e.}}
\def\eg{{\it e.g.}}
\def\be{\begin{equation}}
\def\ee{\end{equation}}
\def\bseq{\begin{subequations}}
\def\eseq{\end{subequations}}

\def\bea{\begin{eqnarray}}
\def\eea{\end{eqnarray}}
\newcommand\bbone{\ensuremath{\mathbbm{1}}}
\newcommand{\ul}{\underline}
\def\bseq{\begin{subequations}}
\def\eseq{\end{subequations}}

\numberwithin{equation}{section} 
\usepackage[]{graphicx}

\def\d {{\rm d}}

\def\X  {{X}}
\def\Y  {{Y}}

\def\0         {{\it 0}}
\def\1         {{\it 1}}
\def\2         {{\it 2}}
\def\3         {{\it 3}}
\def\4         {{\it 4}}
\def\5         {{\it 5}}
\def\6         {{\it 6}}
\def\7         {{\it 7}}
\def\8         {{\it 8}}
\def\9         {{\it 9}}

\def\cala         {{\cal A}}

\def\cale         {{\cal E}}
\def\calf         {{\cal F}}
\def\calg         {{\cal G}}

\def\calk         {{\cal K}}
\def\call         {{\cal L}}
\def\calm         {{\cal M}}
\def\caln         {{\cal N}}
\def\calo         {{\cal O}}

\def\calr         {{\cal R}}
\def\cals         {{\cal S}}

\def\calu         {{\cal U}}

\def\R         {{\rm R}}
\def\L          {{\rm L}}

\def\RH         { right-handed }
\def\LH          { left-handed }

\def\rhdot         {r.h.}
\def\lhdot          {l.h.}

\def\del          {\partial}
\def\delbar       {\bar\partial}
\def\ii           {{\rm i}}

\def\Re           {{\rm Re\hskip0.1em}}
\def\Im           {{\rm Im\hskip0.1em}}

\def\sqr#1#2{{\vcenter{\vbox{\hrule height.#2pt
 \hbox{\vrule width.#2pt height#1pt \kern#1pt \vrule width.#2pt}\hrule
 height.#2pt}}}}

\def\d{\text{d}}

\def\slashchar#1{\setbox0=\hbox{$#1$}           
\dimen0=\wd0                                 
\setbox1=\hbox{/} \dimen1=\wd1               
\ifdim\dimen0>\dimen1                        
\rlap{\hbox to \dimen0{\hfil/\hfil}}      
#1                                        
\else                                        
\rlap{\hbox to \dimen1{\hfil$#1$\hfil}}   
/                                         
\fi}

\begin{document}
\font\cmss=cmss10 \font\cmsss=cmss10 at 7pt

\vskip -0.5cm

\vskip .7 cm

\begin{flushright}{\scriptsize LMU-ASC 31/11 \\  \scriptsize  ROM2F/2011/09}
\end{flushright}
\hfill
\vspace{18pt}
\begin{center}
{\Large \textbf{Magnetized E3-brane instantons \\ in F-theory}}
\end{center}

\vspace{6pt}
\begin{center}
{\textsl{ Massimo Bianchi$\,^\heartsuit$\footnote{\scriptsize \tt massimo.bianchi@roma2.infn.it}, Andr\'es Collinucci$\,^\diamondsuit$\footnote{\scriptsize \tt andres.collinucci@physik.uni-muenchen.de} \& Luca Martucci$\,^\heartsuit$\footnote{\scriptsize \tt luca.martucci@roma2.infn.it}}}

\vspace{1cm}
$^\heartsuit$\textit{\small I.N.F.N. Sezione di Roma ``TorVergata'' \&\\  Dipartimento di Fisica, Universit\`a di Roma ``TorVergata", \\
Via della Ricerca ScientiÞca, 00133 Roma, Italy }\\  \vspace{6pt}
$^\diamondsuit$\textit{\small Arnold Sommerfeld Center for Theoretical Physics,\\ LMU M\"unchen, Theresienstra\ss e 37, D-80333 M\"unchen, Germany}\\  \vspace{6pt}
\end{center}


\vspace{12pt}

\begin{center}
\textbf{Abstract}

\end{center}

\vspace{4pt} {\small
\noindent We discuss E3-brane instantons  in $\caln=1$ F-theory  compactifications to four dimensions and clarify the structure of
E3-E3 zero modes for general world-volume fluxes. We consistently incorporate SL$(2,\mathbb{Z})$ monodromies and highlight
the relation between F-theory and perturbative IIB results. We explicitly show that world-volume fluxes can lift certain fermionic
zero-modes, whose presence would prevent the generation of
non-perturbative superpotential terms, and we discuss in detail the
geometric interpretation of the zero-mode lifting mechanism.
We provide a IIB derivation of the index for generation of superpotential terms and of its modification to include world-volume fluxes,
which reproduces and generalizes available results.
We apply our general analysis to the explicit, though very
simple, example of compactification on $\mathbb{P}^3$ and its orientifold weak-coupling
limit. In particular, we provide an example in which a non-rigid divisor
with fluxes contributes to the superpotential.
\noindent }

\vspace{1cm}


\thispagestyle{empty}


\newpage

\setcounter{footnote}{0}

\tableofcontents

\newpage


\section{Introduction and summary of the results}

F-theory provides a large class of non-perturbative vacua of Type IIB
whereby the complexified axio-dilaton develops a non trivial profile due
to the presence of 7-branes \cite{Vafa:1996xn, Morrison:1996na,
Morrison:1996pp}. In recent times a resurgence
of interest in this class of models has been triggered by the observation
that local configurations of 7-branes can accommodate interesting aspects
of particle physics phenomenology beyond the standard model, such as gauge
coupling unification and textures in the Yukawa couplings, starting with
\cite{donagiF, Beasley:2008dc, Beasley:2008kw} -- see \eg\ \cite{heckman, timoreview} for reviews and  more complete lists of references.

Although it is still unclear which local F-theory solutions
admit a global embedding, non-perturbative effects generated by Euclidean
D3-branes (E3-branes) are widely recognized as crucial ingredients in
IIB/F-theory model building, see for instance \cite{Finstantons, marsano1, Marchesano:2009rz, Cvetic:2009ah, Cvetic:2010ky, grimm2011}.
Most of the applications involving M5/E3 non-perturbative effects are built on the results  of \cite{witten96}, which in turn relies on the study of the fermionic zero modes
derived from the standard Dirac action for a Euclidean M5-brane, dual to the E3-brane.
However, even in the absence of background fluxes, world-volume fluxes are known to deform the standard Dirac action on D-branes \cite{Ddirac}  and M5-branes \cite{dima}.  Since the effect of world-volume fluxes on the fermionic zero-mode structure is not  taken into account in \cite{witten96}, understanding this and related issues remains an open problem. This constitutes the main motivation of the present investigation, that aims at tackling
the effect of world-volume fluxes on E3-brane instantons in general terms. Henceforth we will dub these as {\em magnetized} E3-branes.

This is not just an academic problem. On the one hand, consistency actually requires that for every four-cycle wrapped by an E3-brane, one must sum over the infinite
(discrete) family of possible world-volume fluxes. On the other hand,
fluxes can in fact have important physical effects. For instance, there is evidence that they can modify the zero-mode counting \cite{timo2007} and thus
the nature of E3-brane instanton corrections to the effective theory. In particular \cite{grimm2011},  fluxes can play a key role in alleviating the tension between chirality and moduli stabilization \cite{ralphchiral}.

While \cite{witten96} and many other papers on F-theory are largely based on  the dual
M-theory point of view, that has of course several advantages, in this paper we will mostly work in the IIB picture.
One practical motivation behind this choice is that it allows
one to use the (Wick-rotated) well-understood effective action for D3-branes, hence avoiding the M5-brane effective action and in particular its subtle chiral three-form flux.
Furthermore, our study of the effect of world-volume fluxes on E3-branes will have as byproduct
a better comprehension of the relation between  the IIB and M-theory pictures even in absence of fluxes -- see \cite{Cvetic:2009ah, andres, donagi2010, Cvetic:2010ky} for previous works in this direction.
An improved understanding of the IIB/M-theory dictionary can be important since
the IIB approach   has the unquestionable advantage of admitting --
in some cases -- local or global weak coupling descriptions, wherein one can use  perturbative string theory techniques, not available on the M-theory side.  Indeed, in the past few years, a lot of  progress has been achieved in the
understanding of non-perturbative effects generated by unoriented D-brane instantons using their open string description \cite{Billo:2002hm, Billo:2006jm}\footnote{See also \cite{Blumenhagen:2006xt, Ibanez:2006da, Florea:2006si, Bianchi:2007fx, Argurio:2007qk, Argurio:2007vqa, Bianchi:2007wy} for early work and \cite{ralphreview, Bianchi:2009ij} for recent reviews.}.

Although in this paper we will not consider background type IIB  three-form fluxes (or, dually, background M-theory  four-form  fluxes),
the understanding of E3-branes in their presence actually constitutes
another motivation for our work. Indeed, background fluxes  play a crucial role in several applications
and their effect on M5/E3 brane instantons in F-theory/IIB orientifold backgrounds  has been already considered in \cite{Gorlich:2004qm,saulinaM5, kallosh1, kallosh2, Lust:2005cu, tsimpis07, timo2007, fernando, uranga08}, see also \cite{Billo:2008sp, Billo:2008pg} for studies on the effect of bulk fluxes on D-brane instantons by using the string world-sheet techniques. However, the papers \cite{saulinaM5, kallosh1, kallosh2, tsimpis07} work within the simplifying assumption that the world-volume flux vanishes and \cite{timo2007,fernando, uranga08} provide just partial results on its possible effects.
In this context, although  legitimate in some cases,
the assumption that there is no world-volume flux constitutes a very non-generic condition,
since the world-volume Bianchi identities relate world-volume and background fluxes.
Take for instance a IIB three-form flux $H_{\it 3}$. The world-volume Bianchi identity reads $\d\calf=-\iota^*H_{\it 3}$,
where $\iota^*H_{\it 3}$ denotes the pull-back of $H_{\it 3}$ onto the  E3-brane world-volume. Clearly,
if $H_{\it 3}\neq 0$, then generically the world-volume flux $\calf$ will be non vanishing as well.
 Hence, a proper understanding of the effect of world-volume fluxes alone on E3/M5-branes
 constitutes an important step that necessarily precedes  a consistent incorporation of background fluxes.


\bigskip

Let us summarize our findings:

\begin{itemize}

\item
We clarify the structure of E3-E3 zero modes for general world-volume
fluxes and propose a modification of the M5-brane index \cite{wittenM5} that reproduces and generalizes all
available results.

\item
We consistently incorporate SL$(2,\mathbb{Z})$ non-trivial monodromies and
highlight the relation between F-theory and perturbative IIB results,
that was not so manifest even in the flux-less case \cite{Cvetic:2009ah,andres,donagi2010}.

\item
We explicitly show that world-volume fluxes can lift certain fermionic zero-modes.
We discuss in detail the associated geometrical interpretation of the
flux-induced fermionic zero-mode lifting mechanism.
We eventually apply our general analysis to the explicit, though very
simple, example with $\mathbb{P}^3$ as compactification space and its orientifold weak-coupling limit.

\item
In particular, we discuss a concrete example that provides evidence that non-rigid divisors can generically
contribute to the superpotential. This drastically broadens the
possibilities for non-perturbatively generated superpotential terms. For
instance,  once properly combined with bulk fluxes,  E3-brane instantons with world-volume fluxes can lead to
significant improvement of the moduli-lifting problem.

\end{itemize}

The plan of the paper is as follows. In Section \ref{sec:Ftheory}, we will
briefly review some basic properties of F-theory vacua, including
monodromy, S-duality and, whenever possible, orientifold limit.
We then pass to discuss in Section \ref{sec:fluxlessE3} the fermionic parts
of the E3 action and the index counting the fermionic zero-modes and
present a very useful though simple working example.
Section \ref{sec:fluxE3} is devoted to study the effect of turning on
world-volume fluxes. In Section \ref{sec:fluxexample} we present an
explicit example whereby a non-rigid divisor
with fluxes contributes to the superpotential.

We have collected some useful formulae on holomorphic line bundles and
properties of the extrinsic curvature in two Appendices.

\bigskip

\noindent {\em Note added}: While this paper was being typed a couple of interesting papers appeared \cite{Marsano:2011nn, cvetic2011} that discuss instantons in F-theory and provide
complementary results to the ones in the present work.

\section{A short review on F-theory vacua}
\label{sec:Ftheory}

The aim of this section is to review the structure of F-theory backgrounds. We will emphasize the IIB viewpoint, that is somewhat less `standard', being
often rephrased in terms of the dual M-theory description. The material covered in this section is of course not new at all, but we have decided to include it for self-completeness of the paper, in order to fix notation, definitions and main properties of these vacua and facilitate the reading of the subsequent sections.  For more details, see for instance the recent reviews \cite{denef, timoreview}.

F-theory vacua are type IIB backgrounds that include full back-reaction to 7-brane induced fluxes.
The 7-branes can be either D7-branes or more general $(p,q)$ 7-branes,
obtained by acting on a D7-brane -- \ie\ a (1,0) 7-brane -- by an SL(2,$\mathbb{Z}$) duality transformation.
A D7-brane sources one unit of Ramond-Ramond (RR) flux $F_{ \1 }=\d C_{\it 0}$ and is then characterized by a monodromy $\tau\rightarrow \tau+1$ of the axion-dilaton $\tau:=C_{ \0 }+\ii e^{-\phi}$ on a closed loop linking the D7-brane. Analogously, a $(p,q)$ 7-brane is characterized by a more general  SL(2,$\mathbb{Z}$) monodromy of $\tau$.

Local and global tadpole constraints strongly restrict the consistent configurations
and often require the presence of mutually non-local 7-branes, \ie\ 7-branes that cannot contemporarily be seen as a set of
D7-branes or an SL(2,$\mathbb{Z}$) transform thereof. In other words, these 7-brane configurations are intrinsically non-perturbative, in the sense
that the axion-dilaton $\tau$ generically undergoes  SL(2,$\mathbb{Z})$ duality transformations when going from one patch of the internal manifold to another. A simple example of such a configuration is provided by the O7-planes of perturbative string theory, more appropriately described in F-theory as a bound states of two mutually non-local 7-branes \cite{sen1,sen2}.

In this paper we are interested in (minimally) supersymmetric  F-theory compactification to four-dimensions.
They are dual to M-theory compactifications to three-dimensions on an elliptically fibered Calabi-Yau four-fold, in which the complex structure
 of the elliptic fiber corresponds to the axion-dilaton $\tau$. This dual description nicely geometrizes the non-trivial features
 of the F-theory backgrounds and turns out to be very convenient to study several aspects that are harder to handle within a direct IIB framework. However, the direct type IIB viewpoint can be important for other purposes, as we will show in this paper.

Hence,  in sections \ref{sec:gen}, \ref{sec:killing} and \ref{sec:O7IIB} we will  describe some basic features of the F-theory backgrounds within the purely IIB description, without making any use of the dual M-theory picture. The latter will be considered only in subsection \ref{sec:Mtheory}. We will also try to make very explicit the relation between the two descriptions. This will be important in the subsequent sections. In section \ref{sec:example} we provide a simple example of F-theory compactification and its weak coupling orientifold limit. This example will constitute the playground on which we will apply  the subsequent general discussions.

\subsection{Generalities on F-theory compactifications}
\label{sec:gen}

F-theory backgrounds are more conveniently described in the Einstein frame. Focusing on compactifications to flat four-dimensional Minkowski space and demanding supersymmetry, the ten-dimensional space is the direct product $\mathbb{R}^{1,3}\times \X$, where $\X$ is a complex three-fold, \ie\ a real six-dimensional manifold, that is also complex and K\"ahler.  The Einstein frame ten-dimensional metric splits as
 \be
\d s^2_{10}=\d x^\mu\d x_\mu+\d s^2_{\X}
\ee
Here $\d s^2_{\X}=g_{mn}\d y^m\d y^n$ is the K\"ahler metric on ${\X}$. In this paper we denote the complex structure by $I=I^m{}_n\d y^n\otimes \partial_m$ and the associated K\"ahler form by $J=\frac12 J_{mn}\d y^m\wedge \d y^n$, where $J_{mn}=g_{mk}I^k{}_n$.
One can also  include a non-trivial warping, that can be sourced by possible D3-brane charge and three-form flux present on $\X$, but in the present paper we will focus on the cases in which such warping can be consistently considered constant.

As recalled at the beginning of this section, the physically non-trivial feature characterizing the F-theory backgrounds is provided by the presence of 7-branes. In the present setting, they fill $\mathbb{R}^{1,3}$ and wrap internal four-cycles that are required to be holomorphically
embedded by supersymmetry. Hence, by adopting the algebraic geometry terminology, the 7-branes wrap {\em divisors} of the internal space.
The presence of 7-branes is signaled by the presence of a non-trivial axion-dilaton $\tau$, that supersymmetry requires  to depend holomorphically on the internal coordinates
\be\label{holtau}
\delbar\tau=0
\ee
where $\bar\partial := \d \bar z^{\bar{ I}}\wedge \partial/\partial \bar z^{\bar{  I}}$, where $(z^{  I},\bar z^{\bar{  I}})$ are complex coordinates on $\X$.

In general, when passing from one local patch on $\X$ to another, the axion-dilaton is allowed to experience non-trivial SL(2,$\mathbb{Z}$) duality transformations
\be\label{dualtau}
\tau \quad \rightarrow\quad \tau^\prime=\frac{a\tau+b}{c\tau+d}
\ee
with $a,b,c,d\in\mathbb{Z}$ and $ad-bc=1$. In particular, non-trivial monodromies signal the presence of 7-branes. A D7-branes is associated to a so-called $T$-monodromy
\be
M_{\rm D7}\equiv M_{1,0}=\left(\begin{array}{cc}   1  & 1 \\ 0 & 1 \end{array}\right)\equiv T
\ee
This implies that close to a D7-brane $\tau$ takes the form $\tau\simeq \frac{1}{2\pi \ii}\log (z-z_{\rm D7})$, where $z$ denotes some local coordinate transverse to the D7-brane. Indeed, close to a D7-brane $\tau$ must satisfy the second-order equation $\del\delbar\tau=-\delta^{\it 2}({\rm D7})$. More generically, the monodromy around a $(p,q)$ 7-brane is given by
\be
M_{p,q}=\left(\begin{array}{cc}   1-pq & p^2 \\ -q^2 & 1+pq \end{array}\right)
\ee

The non-trivial axion-dilaton contributes to the energy-momentum tensor. Hence the internal metric is not Ricci-flat and indeed supersymmetry implies the following Einstein equations
\be\label{ricci}
R^{\X}_{{  I}\bar{  J}}= \nabla_{  I}\bar\nabla_{\bar{  J}}\phi
\ee

It is useful to introduce the following composite one-form
\be \label{Qconnection}
Q_{\it 1}=\frac{\ii}{2}\frac{\d(\tau+\bar\tau)}{\tau-\bar\tau}=\frac12\, e^\phi \, F_{\it 1}=\frac{\ii}2(\delbar\phi-\del\phi)
\ee
where, in the last equality, we have made use of (\ref{holtau}). Now, $Q_{\it 1}$ can be seen as a connection for a U(1) bundle, that we call U(1)$_Q$. It is defined as follows: If the background undergoes an SL(2,$\mathbb{Z})$ duality transformation
\be
\left(\begin{array}{cc} a & b \\ c & d\end{array}\right)
\ee
when going from one patch to another, the corresponding U(1)$_Q$ transition function is given by $e^{\ii \arg(c\tau+d)}$. This U(1)$_Q$ bundle will play a crucial role in the following, in that several fields can be seen to transform as sections of the associated complex line $L_Q$ and powers thereof under SL(2,$\mathbb{Z}$). Hence, $Q_{\it 1}$ can be used to construct   SL(2,$\mathbb{Z}$)-covariant derivatives,  that allow one to obtain manifestly SL(2,$\mathbb{Z}$)-covariant quantities. This will be very important in the following discussions.

From (\ref{Qconnection}), one can easily compute the curvature associated with $Q_{\it 1}$:
\be\label{Qcurv}
\calr_Q=\d Q= \ii\del\delbar \phi
\ee
Then, by (\ref{ricci}) it is immediate to check that it is equal to the Ricci form  $\calr_{\X}:=R^{\X}_{{  I}\bar{  J}}\,\d z^{  I}\wedge \d\bar z^{\bar{  J}}$:
\be\label{curveq}
\calr_{\X}\equiv \calr_Q
\ee

The curvature $\calr_Q$ computes the first Chern Class of the line bundle $L_Q$, $c_1(L_Q)=\frac{1}{2\pi}\left[\calr_Q\right]$, while the Ricci form computes the first Chern Class  of the tangent bundle of $\X$, $c_1(\X)=\frac{1}{2\pi}\left[\calr_{\X}\right]$. Hence, one can explicitly see that the non-triviality of the line bundle $L_Q$ is directly related to the failure of the Calabi-Yau condition for $\X$. Furthermore, by Yau's theorem, given a certain holomorphic axion-dilaton $\tau$ with associated bundle $L_Q$ on $\X$, the topological condition $c_1(L_Q)=c_1(\X)$ is actually sufficient  for an F-theory metric to exist for every K\"ahler class $[J]\in H^{1,1}(\X)$.

Notice that, by standard arguments in algebraic geometry, since $\calr_Q$ has vanishing (2,0) and (0,2) parts, one can associate to  $L_Q$ a {\rm holomorphic} line bundle $\call_Q$, whose sections $f$ satisfy holomorphic gluing conditions $f\rightarrow (c\tau+d)f$ -- see appendix \ref{sec:bundles} for more details. As we will explicitly see, the possibility to trade $L_Q$ for $\call_Q$ will allow to translate several geometrical quantities in terms of purely holomorphic data.

For instance, it is well known that $c_1(\X)=-c_1(K_{\X})$, where $K_{\X}$ is the canonical bundle on $\X$, \ie\ the holomorphic bundle of $(3,0)$ forms on $\X$. Hence, (\ref{curveq}) implies that $c_1(\call_Q)=-c_1(K_{\X})$ and one can conclude that the holomorphic line bundle $\call_Q$ is isomorphic to the inverse of the canonical bundle on $\X$:
\be\label{KLiso}
\call_Q\simeq  K^{-1}_{\X}
\ee

\subsection{Supersymmetric structures}
\label{sec:killing}

We are interested in minimally supersymmetric F-theory compactifications, \ie\ preserving four-supersymmetries.
The supersymmetry generators are described by the two type IIB Majorana-Weyl  spinors $\epsilon_{1,2}$.
They both have positive chirality $\Gamma_{11}\epsilon_{1,2}=\epsilon_{1,2}$, where $\Gamma_{11}=\Gamma^{\ul{0123456789}}$ is the ten-dimensional chirality operator.\footnote{In this paper, we distinguish flat indices by underlying them.} We work in the Majorana representation in which the charge conjugation matrix is $\Gamma^{\ul 0}$ and all ten-dimensional gamma matrices are real (in Minkowskian signature). Hence, one can take $\epsilon_{1,2}$ to be real. One can group them in a bi-spinor $\epsilon$
\be
\epsilon=\left(\begin{array}{c} \epsilon_1 \\ \epsilon_2  \end{array}\right)
\ee

In order to describe the supersymmetry generators, one can first introduce the following split of the ten-dimensional gamma matrices
\be\label{gammadec}
\Gamma^{\ul{\mu}}=\hat\gamma^{\ul{\mu}}\otimes \bbone\, ,\quad \Gamma^{\ul{m}}=\gamma_5\otimes \gamma^{\ul{m}}
\ee
where $\hat\gamma^{\ul \mu}$ and $\gamma^{\ul{m}}$ are four- and six-dimensional gamma matrices respectively, and furthermore
\be\label{chiralm}
\Gamma_{11}=\gamma_5\otimes\gamma_{7}\, ,\quad \hat\gamma_5=-\ii\hat\gamma^{\ul{0123}}\, ,\quad \gamma_7=\ii\gamma^{\ul{123456}}
\ee
with $\hat\gamma_5$ and $\gamma_7$ the  four- and six-dimensional chirality operators respectively.

In the Einstein frame, the supersymmetry generators
 in double-notation  have the structure
\be\label{spinordec}
\epsilon=\epsilon^\R+\epsilon^\L=\left(\begin{array}{c} \epsilon^\R_1 \\ \epsilon^\R_2  \end{array}\right)
+\left(\begin{array}{c} \epsilon^\L_1 \\ \epsilon^\L_2  \end{array}\right)
 \ee
 where
 \be\label{killingspinor}
\epsilon^\R_1=\ii \epsilon_2^\R=\zeta_{\R }\otimes\eta\quad,\quad  \epsilon_1^\L = -\ii\epsilon_2^\L=\zeta_\L\otimes\eta^*
\ee
with $\zeta_{\R,\L}$ and $\eta$ external and internal chiral spinors: $\hat\gamma_5\zeta_\R =\zeta_\R $, $\hat\gamma_5\zeta_\L=-\zeta_\L$ and $\gamma_7\eta=\eta$. In Minkowskian signature, one has $\zeta_\L=\zeta^*_{\R }$. However, in order to consider Euclidean D-branes, one needs to analytically continue $\zeta_{\R }$ and $\zeta_\L$ to two independent  `holomorphic' 4D spinors of opposite chirality.

Notice that the complex combination $\epsilon_1+\ii\epsilon_2$ transforms with U(1)$_Q$-charge $+1/2$ under the SL(2,$\mathbb{Z}$) duality group. Hence, the same is true for $\eta$, that  can be then considered a section of $S_+\otimes L_Q^{1/2}$ or, more precisely, of the associated Spin$^c_+$ bundle over $\X$, that reduces to $S_+\otimes L_Q^{1/2}$ only if  $S_+$ and $L_Q^{1/2}$ separately exist. Indeed, consistently with the discussion of the previous subsection,  $\eta$ must satisfy the following SL(2,$\mathbb{Z}$)-covariant Killing spinor equations
\be
(\nabla_m-\frac{\ii}{2}\, Q_m)\eta=0
\ee
where $\nabla_m$ is the Levi-Civita spin connection.
In the following we will choose  the normalization condition $\eta^\dagger\eta=1$.

The K\"ahler form $J$ on $\X$ can be constructed from $\eta$ as
\be\label{Jform}
J_{mn}=\ii\eta^\dagger\gamma_{mn}\eta  \quad
\ee
It is closed and in fact covariantly constant since it does not transform under the structure group of $L_Q$.
On the other hand, one can use $\eta$ to construct also a holomorphic (3,0) form $\Omega$ as follows
\be\label{omega}
\Omega_{mnp}=e^{\phi/2}\eta^T\gamma_{mnk}\eta
\ee
Notice that
\be
\Omega\wedge \bar\Omega= \frac{4\ii}{3} e^{\phi}J\wedge J\wedge J
\ee
This explicitly shows that the metric is not Calabi-Yau if the dilaton is not constant.

We would like to emphasize that $\Omega$ should not be considered as a section of the canonical bundle, since one can easily check that it transforms non-trivially under SL(2,$\mathbb{Z}$)-duality:
\be
\Omega\rightarrow (c\tau+d)\Omega
\ee
Hence, $\Omega$ must be rather considered as a section of $K_Q\otimes \call_Q$. Because of the isomorphism (\ref{KLiso}), one can see that $\Omega$ is a holomorphic section of a trivial line-bundle, which guarantees that $\Omega$ is  globally well-defined and never vanishing.

Notice that $\Omega$ is not covariantly constant. On the other hand, according to the general discussion of appendix \ref{sec:bundles}, one can associate to $\Omega$ the $L_Q$-valued three-form $e^{-\phi/2}\Omega$, that is  covariantly constant  under the correspondingly U(1)$_Q$-covariant derivative:
\be
\nabla_m^Q(e^{-\phi/2}\Omega)\equiv (\nabla_m-\ii Q_m)(e^{-\phi/2}\Omega)=0
\ee

 \subsection{Orientifold limit}
 \label{sec:O7IIB}

Backgrounds with O7-planes and D7-branes can be properly described as particular limits of the more general F-theory backgrounds
considered in the previous sections \cite{sen1,sen2}, in which the string coupling $g_s$ can be tuned to be small. In this limit, the O7-plane can be seen as a bound state of two mutually non-local 7-branes separated by a perturbatively invisible distance suppressed by $e^{-1/g_s}$. Neglecting such a non-perturbatively resolution, the O7-plane appears as a singular non-dynamical object characterized by an SL(2,$\mathbb{Z}$) monodromy
\be
M_{\rm O7}=\left(\begin{array}{cc}   -1 & -4 \\ 0 & -1 \end{array}\right)
\ee
When four D7-branes coincide with one O7-plane, there is a net cancellation of the resulting backreaction, with a residual monodromy
\be\label{O7D7monodromy}
M_{\rm O7}M^4_{\rm D7}=\left(\begin{array}{cc}   -1 & 0 \\ 0 & -1 \end{array}\right)
\ee
This monodromy does not act on the axion-dilaton but reverses the sign of the fields transforming as doublets under SL(2,$\mathbb{Z}$). The standard way to describe this effect is to construct a double cover $\tilde \X$ and describe the F-theory space as a $\mathbb{Z}_2$-quotient $\X=\tilde \X/\sigma$, where $\tilde \X$ is a double-cover of $\X$ branched over the locus of the O7-plane, and $\sigma$ is the orientifold involution. At the level of complex structure, $\tilde \X$ is a Calabi-Yau, in the sense that  the holomorphic (3,0) form $\Omega$ defined in (\ref{omega}) pulls-back to a globally defined nowhere-vanishing section of the canonical bundle of $\tilde \X$. When all 7-branes organize in groups of O7+4 D7, then $\tau$ is globally constant and $\Omega$ is covariantly constant too, so that $\tilde \X$ has Ricci-flat Calabi-Yau metric.

One may also relax this limit where the axion-dilaton is constant by allowing a configuration with an O7-plane and one or several D7-branes that do not lie on it. Then, the only SL(2,$\mathbb{Z}$) monodromies will be those that shift $\tau$ by real constants, and the sign reversing orientifold action on doublets. In such a limit, $g_s$ can be kept small everywhere except for an exponentially small region around the O7-plane. Hence, one can still proceed within perturbative string theory as long as one is aware of this subtlety.

\subsection{Relation with the M-theory viewpoint}
\label{sec:Mtheory}

The backgrounds described in the previous subsections are dual to M-theory compactifications on elliptically fibered Calabi-Yau four-folds $\Y$ -- see for instance \cite{denef} for a detailed discussion of this duality. In this section we would like to provide the concrete prescription on how to construct $\Y$ starting from the type IIB data.

Given a holomorphic axion-dilaton on a K\"ahler three-fold  $\X$, one defines an elliptic fibration over it by first creating a fiber bundle over it, whose fibers are the weighted projective space $\mathbb{CP}^2_{2 3 1}$ with homogeneous coordinates $[x:y:z]$. One way to make this bundle non-trivial over $\X$ is by allowing the projective coordinates to transform as sections of line bundles $\call_x$, $\call_y$ and $\call_z$ on $\X$ under transitions between patches in $\X$.  In order to make the elliptically fibered four-fold, we would like to impose fiber by fiber the Weierstrass equation:
\be\label{weierstrass}
y^2 = x^3+f\,x\,z^4+g\,z^6
\ee
that cuts out an elliptic curve inside the weighted projective space. In particular, the complex structure of the elliptic fiber must coincide with the axion-dilaton $\tau$ on $\X$.

On the other hand, it is a standard fact that elliptic curves with such a Weierstrass representation have very tractable modular properties. More precisely, it is known that, under an SL(2,$\mathbb{Z}$) transformation, $f$ and $g$ transform as modular forms of weight 4 and 6 respectiverly, i.e.\
\be\label{modforms}
f\rightarrow (c\tau+d)^4 f\quad,\quad g\rightarrow (c\tau+d)^6 g
\ee
By identifying the modular transformations of the elliptic fiber with the S-duality group of IIB string theory, one can clearly see that according to the definition in \ref{sec:gen} of the line bundle $\call_Q$, $f$ and $g$ can be regarded as holomorphic sections of $\call^4_Q$ and $\call^6_Q$:
\be
f \in \Gamma(\call_Q^{4}) \quad {\rm and} \quad g \in \Gamma(\call_Q^{6})
\ee
Then, by requiring consistency of (\ref{weierstrass}), one can see that $\call_x= \call_z^2\otimes \call^2_Q$ and $\call_y= \call_z^3\otimes \call^3_Q$. At this point, one has some freedom in the choice $\call_z$. One can for instance take $\call_z=\calo_{\X}$ (\ie\ the trivial line bundle on $\X$), and then  $\call_x= \call^2_Q$ and $\call_y= \call^3_Q$. This choice facilitates the description of the open patch where $z \neq 0$, by allowing one to gauge fix $z \rightarrow 1$, and recovering the familiar Weierstrass equation.
 Another useful possibility is $\call_z=\call_Q^{-1}$, for which $\call_x= \call_y= \calo_{\X}$. Since  $\X$ is itself holomorphically embedded in $\Y$ as the divisor $z=0$, this choice is useful because it easily allows us to restrict integrals on $\Y$ to integrals on $\X$.

In section \ref{sec:gen} we have also shown that supersymmetry constrains this line bundle to be isomorphic to the anti-canonical bundle of the three-fold, $\call_Q \simeq K_{B}^{-1}$. Via standard adjunction formulae, one can compute the first Chern class of the F-theory elliptically fibered four-fold $\Y$ to be
\be
c_1(\Y) = c_1(\X) - c_1(\call_Q)
\ee
Hence, since $c_1(\X) = - c_1(K_{\X})$, the supersymmetry condition  $\call_Q \simeq K_{B}^{-1}$ is equivalent to imposing the Calabi-Yau condition on $\Y$.

\vskip 3mm

Having established the basics about the M-theory perspective as well as the IIB supergravity point of view, let us connect it to the perturbative IIB string theory description via Sen's limit \cite{sen1, sen2}. Given a choice for $f$ and $g$, the corresponding axion-dilaton is determined by the following relation:
\be
j(\tau) = \frac{4\,(24\,f)^3}{4\,f^3+27\,g^2} \quad {\rm where} \quad j(\tau) = e^{-2\pi \ii \tau}+744+ \calo(e^{2 \pi \ii \tau})
\ee
is Klein's modular function. Let us reparametrize $f$ and $g$ as follows:
\begin{eqnarray}
f &=& -3\,h^2+\epsilon\,\eta\\
g &=& -2\,h^3+\epsilon\,h\,\eta-\epsilon^2\,\chi/12
\end{eqnarray}
where $h, \eta,$ and $\chi$ are sections of $\call_Q^2, \call_Q^4,$ and $\call_Q^6$, respectively.
Then, one finds that $g_s \sim (\log\epsilon)^{-1}$ everywhere except at the locus $h=0$. A monodromy analysis reveals that the $(p,q)$ branes have rearranged into the following perturbative configuration:
\begin{eqnarray}
\text{ O7-plane at}:&&\quad h=0 \label{eqO7}\\
\text{D7-brane at}:&& \quad \eta^2-h\,\chi=0 \label{eqD7}
\end{eqnarray}
Away from this limit, however, the total $C_{ \0 }$-tadpole cancelling 7-brane configuration becomes one recombined object wrapping the following divisor:
\be
4\,f^3+27\,g^2=0
\ee
of class $\call_Q^{12}$.

In the non-backreacted probe approximation, where $g_s \rightarrow 0$, we expect the internal space to be Ricci flat, i.e. a Calabi-Yau three-fold. Sen's limit also allows one to recover the internal Calabi-Yau three-fold. It appears as a double-cover of $\X$ branched over the O7-plane locus $h=0$. It is defined by tagging a new coordinate $\xi$ on $\X$, such that $\xi$ transforms as a section of $\call_Q$, and imposing the equation:
\be\label{doublecover}
\xi^2-h=0
\ee
Once we make the choice $\call_Q \equiv K_{\X}^{-1}$, this new space $\tilde \X$ will be guaranteed to be Calabi-Yau.

%

\subsection{A working example}
\label{sec:example}

In order the make the generalities of the previous section more palatable, we will introduce a simple working example. Let the internal three-fold  be $\X = \mathbb{P}^3$, with homogeneous coordinates $[z_1: \ldots : z_4]$.

In general, a section of a holomorphic line bundle vanishes along a complex codimension one holomorphic submanifold, \ie\ a divisor $D$. The line bundle can be then denoted by $\calo(D)$. It turns out that its Poincar\'e dual $[D]$ coincides with the first Chern class of the bundle. On $\mathbb{P}^3$ one can define the so-called \emph{hyperplane bundle}, the line bundle  such that  (linear combinations of) $z_I$ are  sections thereof. The associated divisor is the so-called hyperplane divisor $H$ and the hyperplane bundle is indicated with $\calo_{\mathbb{P}^3}(1)\equiv \calo_{\mathbb{P}^3}(H)$. Analogously, any homogenous polynomial $P^{(n)}(z_1,\ldots,z_4)$ of degree $n$ is a section of $\calo_{\mathbb{P}^3}(n)\equiv \calo_{\mathbb{P}^3}(n\,H)$. Hence, one has the following useful identities
\be
{\rm PD}(\{P^{(n)}=0\}) = c_1(\calo_{\mathbb{P}^3}(n)) = n\,[H]
\ee

The canonical bundle of $\mathbb{P}^3$  is $K_{\mathbb{P}^3} = \calo_{\mathbb{P}^3}(-4)$. Therefore, if $\mathbb{P}^3$ is chosen as F-theory compactification space $\X$,  the supersymmetry of the background requires that we set $\call_Q = \calo_{\mathbb{P}^3}(4)$. From these data, one can easily define the corresponding F-theory Calabi-Yau four-fold $\Y$ as a hypersurface in the following ambient space:
\be\label{cp3example}
\begin{array}{ccccccc|c}z_1&z_2&z_3&z_4&x&y&z&\textrm{eq.}\eqref{weierstrass}\\ \hline
1&1&1&1&0&0&-4&0\\
0&0&0&0&2&3&1&6\end{array}\begin{array}{r}\\ \\ \end{array}
\ee
That is, in terms of the general discussion of section \ref{sec:Mtheory}, one has made the choice $\call_x=\call_y=\calo_{\mathbb{P}^3}$ and $\call_z=\call_Q^{-1}=\calo_{\mathbb{P}^3}(-4)$.

This space is a generalization of a weighted projective space, whereby one takes the quotient w.r.t. two rescalings as follows:
\be
(z_1, \ldots, z_4, x, y, z) \sim (\lambda\,z_1, \ldots, \lambda\,z_4, \mu^2\,x, \mu^3\,y,\,\lambda^{-4}\,\mu\, z)\,, \quad \text{for} \quad \lambda, \mu \in \mathbb{C}^*
\ee
The other necessary data to define this space are the so-called exceptional sets that are excluded from the space, generalizing the usual exclusion of the origin in a projective space. In the present case, there are two sets of coordinates that are forbidden from vanishing simultaneously:
\be
(z_1\,, \ldots \,, z_4) \neq (0, \ldots, 0))\,; \quad {\rm and} \quad (x\,, y\,,z) \neq (0,0,0)
\ee

The Calabi-Yau hypersurface $\Y$ is given by the Weierstrass equation (\ref{weierstrass}),  with $f$ and $g$ sections of  $\calo_{\mathbb{P}^3}(16)$ and $\calo_{\mathbb{P}^3}(24)$ respectively, that is homogeneous of bi-degree $(0,6)$ with respect to the two projective rescalings.

Note that one can easily construct the Calabi-Yau three-fold  $\tilde \X$ for the corresponding perturbative string theory description of the example under consideration in Sen's limit. According to the general discussion of section \ref{sec:Mtheory}, $\tilde \X$ is defined by (\ref{doublecover}), with $h$ section of $\calo_{\mathbb{P}^3}(8)$, that is
\be\label{simpleCY}
\xi^2=h(z_1,\ldots,z_4)
\ee
with $h(z_1,\ldots,z_4)$ a degree 8 homogenous polynomial. Clearly,  (\ref{simpleCY}) can be seen  as a degree eight hypersurface on the weighted projective four-fold $\mathbb{P}^4_{1 1 1 1 4}$ with homogeneous coordinates $[z_1, \ldots, z_4, \xi]$. This Calabi-Yau three-fold is denoted by $\mathbb{P}^4_{1 1 1 1 4}[8]$. On this space, the orientifold involution acts on the target space by sending $\xi \rightarrow -\xi$.


\section{E3-instantons without fluxes: the IIB perspective }
\label{sec:fluxlessE3}

Let us come back to our main motivation, namely the study of E3-instantons in F-theory backgrounds.
In this section we first focus on the case in which there are no world-volume fluxes on the E3-brane, \ie
\be\label{fluxless}
\calf:= 2\pi\alpha^\prime F_{\rm E3}-\iota^*B_{\it 2}=0
\ee
This is  the case considered by Witten in his seminal paper \cite{witten96} and by most of the subsequent papers on this subject.

The approach of \cite{witten96} starts from the M-theory viewpoint and identifies the conditions under which an M5-brane instanton can contribute to the superpotential. Here we would like to retrace  Witten's  procedure working directly in IIB. In this way we will set the basis for our discussion on the inclusion of world-volume fluxes, that will be considered in section \ref{sec:fluxE3}. As a byproduct, we will obtain a clear physical picture on how the M-theory results of \cite{witten96} should be interpreted from the IIB viewpoint. This can be useful in other developments based on the IIB picture, in which one can take advantage of perturbative string theory techniques. See for instance \cite{andres, donagi2010, cvetic2011} for other papers exploring this perspective.

Consider an E3-brane instanton wrapping a four-cycle $D$ in the internal space $\X$.  In order to preserve supersymmetry the four-cycle must be holomorphically embedded or, in other words, must be an effective divisor. Hence, the on-shell E3-action is just given by the value of the complexified K\"ahler modulus
 \be\label{defclosed}
T_{\rm E3}:=\frac{2\pi}{\ell_s^4}\int_D\left(\frac12\, \,J\wedge J+\ii C_{\it 4}\right)
\ee
where $\ell_s:= 2\pi\sqrt{\alpha'}$, and then the possible correction to the superpotential  looks like
\be\label{Wstructure}
W_{\rm np}=\cala(\ldots)\, e^{-T_{\rm E3}}
\ee
where $\cala(\ldots)$ can depend generically on other moduli in the compactification \cite{wittenM5, ganor}, see also \cite{grimmNP}.\footnote{ If in addition 7-brane U(1) fluxes are turned on,  $\cala(\ldots)$ acquires a dependence on gauge-invariant combinations of the charged massless chiral multiplets up to anomalous U(1)'s under which $T_{\rm E3}$ shifts  \cite{Blumenhagen:2006xt, Florea:2006si, Bianchi:2007fx, Argurio:2007vqa, Argurio:2007qk, Bianchi:2007wy}. }

In order to understand the precise structure of $W_{\rm np}$  one should in principle study the complete instantonic path integral. This problem  may be attacked by either using Green-Schwarz-like effective actions, as in \cite{Becker:1995kb,witten96,harveymoore,witten1,witten2,witten3}, or  by microscopic string theory techniques, as  in  \cite{Blumenhagen:2006xt, Florea:2006si, Bianchi:2007fx, Argurio:2007vqa, Argurio:2007qk, Bianchi:2007wy}. In particular, in order to understand whether an E3-brane can contribute to the superpotential, one has to study the structure of the fermionic zero-modes.  Here we are going to focus on the fermions associated to (putative) `open strings' with both ends on the E3-brane, whose dynamics is described by a Green-Schwarz-like effective action that  incorporates  the effect of the non-trivial axion-dilaton in a controlled way. As we will explicitly check, these fermions correspond to the fermions in the dual M5-brane effective action. There could be additional chiral fermions localized at the intersection of the E3-brane with the background 7-branes, associated with (putative) `open strings' connecting the E3 and the 7-branes. In the dual M5-brane, they are incorporated in the chiral (self-dual) three-form and their effect can be reabsorbed in the moduli dependence of the pre-factor $\cala(\ldots)$ in (\ref{Wstructure}) \cite{wittenM5}.

\subsection{Fluxless E3-fermionic action}
\label{sec:fluxlessferm}

In order to study the E3-brane fermionic zero modes, we start from the effective action for the E3 fermions on an F-theory IIB background.
By some general arguments \cite{sadov}, the world-volume fermions should naturally experience a topological twist and then be represented
by world-volume forms. Here we would like to explicitly derive the corresponding topologically twisted theory from the general E3 action. As we will see, a new key role with respect to the more standard topological twist on K\"ahler spaces will be provided by the non-trivial axion-dilaton.
Understanding the role role of the axion-dilaton in this case will be important later, when one will introduce world-volume fluxes.

The fermions on a D-brane in a general supergravity background are more conveniently described in the Green-Schwarz formalism, in which the ordinary bosonic embedding is substituted by an embedding in the ten-dimensional superspace. Hence, the world-volume fermions are described by a pair of ten-dimensional Majorana-Weyl spinors $\theta_1,\theta_2$, that one can combine into the two-component vector
 \be\label{GSfermion}
\Theta=\left(\begin{array}{c} \theta_1 \\
\theta_2\end{array}\right) \ee
The apparent mismatch between bosonic and ferminic degrees of freedom is cured by the presence of a world-volume gauge-symmetry, usually called $\kappa$-symmetry, that acts in the following way
\be
\delta_\kappa\Theta=(\bbone +\Gamma_{\rm E3})\kappa
\ee
where $\kappa=(\kappa_1,\kappa_2)$, with $\kappa_{1,2}$ two arbitrary Majorana-Weyl spinors. In the lowest order expansion in the fermionic fields, the operator $\Gamma_{\rm E3}$ depends just on the bosonic world-volume degrees of freedom and one natural way to remove these redundant degrees of freedom is to impose the $\kappa$-fixing condition
\be\label{genkappafix}
(\bbone +\Gamma_{\rm E3})\Theta=0
\ee
In the present case, since for the moment one is assuming (\ref{fluxless}),  $\Gamma_{\rm E3}$ is given by
 \be\label{gammaE3}
\Gamma_{\rm E3}=\left(\begin{array}{cc} 0 & \hat\gamma_{\rm E3}
\\ \hat\gamma^{-1}_{\rm E3} & 0\end{array}\right)\qquad\qquad\text{with}\qquad \hat\gamma_{\rm
E3}=-\frac{\ii}{4!}\, \frac{\epsilon^{a_1\ldots a_4}}{\sqrt{\det
h}}\,\Gamma_{a_1\ldots a_4}
\ee
We have introduced world-volume coordinates $\sigma^a$, $a=1,\ldots, 4$, $h\equiv g|_D$ denotes the pull-back of the bulk-metric onto the E3 world-volume $D$
\be
\label{pullbmetric}
h_{ab} = g_{mn}(y) \frac{\partial y^m}{\partial\sigma^a} \frac{\partial y^n}{\partial\sigma^b}
\ee
and  $\Gamma_a$ denotes the pull-back of the ten-dimensional gamma-matrices

The explicit $\kappa$-symmetric quadratic fermionic action for D-branes on  general bosonic backgrounds has been worked out, in the string frame, in \cite{Ddirac}. Here one has just to apply the general result of \cite{Ddirac} to the F-theory backgrounds described in detailed in section \ref{sec:Ftheory}.  By taking just a little care of the passage from string to Einstein frame\footnote{Notice that world-volume fermions are rescaled as $\Theta\rightarrow e^{\phi/8}\Theta$ when passing from string to Einstein frame.} and imposing the $\kappa$-fixing condition (\ref{genkappafix}), one obtains the following Green-Schwarz action for an E3-brane wrapping a four-cycle $D\subset B$:
\be\label{thetaction}
S_{\rm F}=\frac{2\pi \ii}{\ell^4_s}\int_{D}
\d^4\sigma\,\sqrt{\det h} \, \,\overline{\Theta}\Gamma^a (\hat \nabla_a
+\frac{\ii}4\, e^\phi F_a\sigma_2)\Theta
\ee
Here $\hat\nabla_a$ is the pull-back of the bulk covariant derivative and  $F_a$ is the pull-back  of the R-R one-form $F_{\it 1}$.
Furthermore $\overline\Theta\equiv \Theta^T\Gamma^{\ul 0}$.

So far, we have not explicitly used the supersymmetry condition on $D$. As already mention, this boils down to the requirement that $D$ should be a holomorphically embedded hypersurface or, in other words, an effective divisor. More specifically, if $D$ is an effective divisor then the E3-brane preserves the two $\epsilon_\R $ supersymmetries as defined in (\ref{spinordec}) and (\ref{killingspinor}), while if $D$ is a negative divisor, that is a anti-holomorphic hypersurface, then the two $\epsilon_\L$ are preserved.

One way to derive these results is by imposing the usual supersymmetry condition
\be\label{generalsusy}
\Gamma_{\rm E3}\epsilon_\R =\epsilon_\R 
\ee
for an effective divisor, or alternatively $\Gamma_{\rm E3}\epsilon_\L=\epsilon_\L$ for a negative divisor. Focusing on E3-branes preserving (\ref{generalsusy}), by using the decompositions (\ref{gammadec}) and (\ref{killingspinor}),  one can more explicitly write (\ref{generalsusy}) as
a condition involving only the internal space:
\be\label{E3susy}
\gamma_{\rm E3}\eta=\ii \eta\quad\quad\text{($\zeta_\R $ preserved)}
\ee
with
\be\label{E3chiral}
\gamma_{\rm
E3}=-\frac{\ii}{4!}\, \frac{\epsilon^{a_1\ldots a_4}}{\sqrt{\det
h}}\,\gamma_{a_1\ldots a_4}
\ee

We are now in a position to introduce a more natural parametrization of the world-volume fermion $\Theta$, that explicitly uses supersymmetry
and automatically solves the $\kappa$-fixing condition (\ref{genkappafix}). Indeed, one can decompose the two components $\theta_1$ and $\theta_2$ as follows. Split $\theta_1=\theta_1^\R+\theta_1^\L$ and
$\theta_2=\theta_2^\R+\theta_2^\L$ and set:
\bea\label{Dfermsplit}
&&\left\{ \begin{array}{l} \theta_1^\R=\frac12\big(\tilde\lambda\otimes \eta+\tilde\psi_a\otimes \gamma^a\eta^*+\frac12\tilde\rho_{ab}\otimes \gamma^{ab}\eta \big)
\\
\theta_2^\R=\frac{\ii}2\big(\tilde\lambda\otimes \eta-\tilde\psi_a\otimes \gamma^a\eta^*+\frac{1}2\tilde\rho_{ab}\otimes \gamma^{ab}\eta\big)

\end{array}  \right. \cr
&& \left\{ \begin{array}{l}
\theta_1^\L=\frac12\big(\lambda\otimes \eta^*+\psi_a\otimes \gamma^a\eta+\frac12\rho_{ab}\otimes \gamma^{ab}\eta^*\big)
 \\
\theta_2^\L=\frac{\ii}{2}\big(\lambda\otimes \eta^*-\psi_a\otimes \gamma^a\eta+\frac{1}2\rho_{ab}\otimes \gamma^{ab}\eta^*\big)
\end{array}  \right.
\eea
Hence, one has traded $\Theta$ for a new sets of world-volume fermionic fields. They are not spinors on $D$ but rather forms, hence explicitly realizing the expected topological twist \cite{sadov}, with purely holomorphic or anti-holomorphic indices. More explicitly, if $s^i$ ($\bar s^{\bar\imath}$) are
(anti-)holomorphic world-volume coordinates, one can write
\be
\tilde\psi=\tilde\psi_i\,\d s^i\, ,\quad \psi=\tilde\psi_{\bar\imath}\,\d \bar s^{\bar\imath}\, , \quad \tilde\rho=\frac12 \rho_{\bar\imath\bar\jmath}\,\d \bar s^{\bar\imath}\wedge \d \bar s^{\bar\jmath}\, ,\quad  \rho=\frac12 \rho_{ij}\,\d s^i\wedge\d s^j
\ee

On the other hand, the new world-volume fermions keep their spinor nature related to four-dimensional external flat space.
Namely, if $S_\pm$ denote the (anti-)chiral spin bundles
associated to  the four flat directions, $\tilde\lambda,\tilde\psi$ and $\tilde\rho$ have  spinorial index in $S_+$, while $\lambda,\psi$ and $\rho$ have  spinorial index in $S_-$.

Finally, since the E3-brane can cover different background patches related by a possible SL(2,$\mathbb{Z}$)-duality transformations,
it is important to understand how the world-volume fermions transform under the duality. By consistency with the superspace formulation,
the Green-Schwarz fermion (\ref{GSfermion}) must transform as as the bulk supersymmetry generator $\epsilon$. Namely, the combinations $\theta_1\pm\ii\theta_2$ must transform
with U(1)$_Q$ charges $\pm1/2$ -- see for instance \cite{kimura}. Since  the internal spinor $\eta$ transforms itself with  U(1)$_Q$ charges $\pm1/2$, it is easy
to recover that $\lambda,\psi$ and $\rho$ are neutral under the SL(2,$\mathbb{Z}$)-duality, while $\tilde\lambda$, $\tilde\psi$ and $\tilde\rho$ transform with
U(1)$_Q$ charges $-1, +1$ and $-1$ respectively.
One can then more synthetically summarize these properties by saying that
\bea\label{Qcharge}
&&\begin{array}{c|c|c} \text{\lhdot\ fermions}& \text{U(1)$_Q$-charge} &\text{associated bundle}\\ \hline
\lambda^\alpha&  0 &  S_-\otimes \Lambda^{0,0}\\
\psi^\alpha & 0 & S_-\otimes \Lambda^{0,1}\\
\rho^\alpha & 0 & S_-\otimes \Lambda^{2,0}
\end{array}\cr  && \\
&&\begin{array}{c|c|c} \text{\rhdot\ fermions} &  \text{U(1)$_Q$-charge} &\text{associated bundle}\\ \hline
\tilde\lambda^{\dot\alpha}&  -1 &S_+\otimes \Lambda^{0,0}\otimes L_Q^{-1}\\
\tilde\psi^{\dot\alpha} & +1 &S_+\otimes \Lambda^{1,0}\otimes L_Q\\
\tilde\rho^{\dot\alpha} & -1 & S_+\otimes \Lambda^{0,2}\otimes L_Q^{-1}\end{array}\nonumber
\eea
where the restriction to $D$ of the bundles defined on $X$ is understood.

By plugging the expansion (\ref{Dfermsplit}) into the action (\ref{thetaction}), after some manipulations, one gets the following effective action for the new fermionic fields
\be\label{twaction2}
S_F=\frac{4\pi\ii }{\ell_s^4}\int_{D}(\psi\wedge
*\del\lambda-\tilde\psi\wedge *\delbar_Q\tilde\lambda-\rho\wedge*\delbar\psi+\tilde\rho\wedge*\del_Q\tilde\psi)
\ee
where $*$ is the Hodge-star computed using the induced metric $h$ and $\del_{Q}$ and $\delbar_Q$ are U(1)$_Q$-covariant
Dolbeault differentials, that is $\del_Q=\del\mp iQ^{1,0}$ and    $\delbar_Q=\delbar\mp iQ^{0,1}$ on fields with  U(1)$_Q$-charge $\pm 1$.

A couple of comments are in order. First of all, the axion-dilaton enter in exactly the right and minimal way
to render the fermionic Lagrangian manifestly invariant under  SL(2,$\mathbb{Z}$)-duality transformations.
The preservation of this important property will be a guiding principle when we will introduce a non-vanishing world-volume flux $\calf\neq 0$.
Second, the action does not depend on the components of background tensorial fields that are  transversal to the E3-brane.
As we will see, this will not any longer be true when $\calf\neq 0$.

\subsection{Zero-modes}
\label{sec:zeromodes}

From the action (\ref{twaction2}) one can easily obtain the following fermionic equations of motion
\bea\label{Feom}
 \del\lambda=0\quad &,& \quad \delbar_Q\tilde\lambda=0 \cr
  \delbar^\dagger\psi=0\quad &,& \quad \del_Q^{\dagger}\tilde\psi=0 \cr
\del^{\dagger}\rho=0\quad &,& \quad \delbar_Q^{\dagger}\tilde\rho=0\\
 \delbar\psi=0
\quad &,& \quad  \del_Q\tilde\psi=0\nonumber
 \eea
where  $\delbar_Q^{\dagger}$ and $\del_Q^{\dagger}$ are the adjoint Dolbeault
operator obtained by using the usual hermitian inner product between forms:
$(\chi_1,\chi_2)=\int \chi_1\wedge *\bar\chi_2$.

Hence, the zero modes are given by the (twisted) {\em harmonic} representatives of the following cohomology groups
\bea\label{sheafcoho}
&&\begin{array}{c|c} \text{\lhdot\ zero modes} &\text{cohomology group}\\ \hline
\lambda^\alpha_{\rm z.m.}&  H_{\del}^{0,0}(D)\simeq  H^0(D,\bar\calo_D)\\
\psi^\alpha_{\rm z.m.} & H_{\delbar}^{0,1}(D)\simeq H^{1}(D,\calo_D)\\
\rho^\alpha_{\rm z.m.} & H_{\del}^{2,0}(D)\simeq  H^2(D,\bar\calo_D)
\end{array}\cr && \\ &&  \begin{array}{c|c} \text{\rhdot\ zero modes} &\text{cohomology group}\\ \hline
\tilde\lambda^{\dot\alpha}_{\rm z.m.}&  H_{\delbar}^{0,0}(D, {L}_Q^{-1})\simeq H^0(D, \call_Q^{-1})\\
\tilde\psi^{\dot\alpha}_{\rm z.m.} &  H_{\del}^{1,0}(D,{L}_Q)\simeq H^{1}(D,\bar\call^{-1}_Q)\\
\tilde\rho^{\dot\alpha}_{\rm z.m.} &  H_{\delbar}^{0,2}(D, {L}_Q^{-1})\simeq H^2(D, \mathcal{L}_Q^{-1})\end{array} \nonumber
\eea
We have already used the fact that, as discussed in section \ref{sec:gen} and more in detail in appendix \ref{sec:bundles}, the $U(1)_Q$ connection $Q$ defines the holomorphic line bundle $\mathcal{L}_Q = K_{\X}^{-1}$, whose holomorphic sections correspond to the $\delbar_Q$-closed section of $L_Q$. This allows to give an interpretation of the fermionic zero modes in terms of sheaf cohomology groups $H^i(D,\calo_D), H^i(D,\call^{-1}_Q)$ and their complex conjugated, as explicitly indicated in (\ref{sheafcoho}).

For later notational convenience, let us define the associated Hodge numbers as follows:
\bea\label{hodgens}
h^{i}(D)&\equiv& {\rm dim}\,   H^{i}(D, \calo_D)={\rm dim}\,   H^{i}(D, \bar\calo_D)\cr
h^{i}_Q (D)&\equiv& {\rm dim}\,   H^{i}(D, \mathcal{L}_Q^{-1})={\rm dim}\,   H^{i}(D, \bar\call_Q^{-1})
\eea
Hence, since each fermion has two four-dimensional spinorial indices, the\LH zero modes are counted by $h^i(D)$, while the\RH zero-modes are counted by $h^i_Q(D)$.

The structure of this fermionic zero-mode spectrum can be interpreted as follows.

First, consider the case in which the restriction of the line bundle $\call_Q$ onto the E3-brane, $\call_Q|_{\rm E3}$, is trivial. Since, as recalled in section \ref{sec:Mtheory}, the seven-brane divisor $D_{7\text{-brane}}$ is associated to a section of  $\call^{12}_Q$, whose restriction on $D$ is also trivial, one is requiring that the intersection two-cycle $D\cap D_{\text{7-brane}}$ is homologically trivial.
In this case, the E3-brane spectrum is transparent to the background 7-branes and is similar to the spectrum for
an E3-brane on a Calabi-Yau three-fold. Namely, the right- and\LH zero-modes are in one-to-one correspondence and have a precise physical interpretation.  First, since $h^0(D)=1$ for a connected $D$, there are  two\LH universal zero-modes $\lambda^\alpha_{\rm z.m.}$, often denoted as $\theta^\alpha$ in the literature, that can be seen as goldstini associated to the supersymmetries $\zeta^\alpha_\L$ broken by the E3-brane. Analogously, in the present case, one has $h^0_Q(D)=1$ too, that  implies that there are two additional\RH zero-modes $\tilde\lambda^{\dot\alpha}$, often denoted as $\bar\tau^{\dot\alpha}$ in the literature -- see for instance \cite{ralphreview} -- associated to the hidden supersymmetry broken by the 7-branes. The other zero-modes have a clear geometrical interpretation. $h^1(D)=h^1_Q(D)$ counts left- and\RH zero-modes that are supersymmetric partners of the Wilson lines, while   $h^2(D)=h^2_Q(D)$ counts left- and\RH zero-modes that are supersymmetric partners to the geometric deformations. The latter statements uses the fact that the restriction to $D$ of the three-fold canonical bundle $K_Q\simeq \call_Q^{-1}$ is trivial too. This implies that $h^2(D)=h^2_Q(D)=\dim H^0(D,N_D)$ through the standard adjunction formula and Serre duality.

Let us now assume that $\call_Q|_{\rm E3}$, is non-trivial, in the sense that the two-cycle $D\cap D_{\text{7-brane}}$ is now homologically non-trivial. In this case, generically, there is a mismatch between left- and\RH zero modes. For instance,
one obviously still has the two universal zero-modes $\lambda^\alpha_{\rm z.m.}\sim \theta^\alpha$, but there could be a different number $2h^0_Q(D)$ of almost-universal zero modes $\tilde\lambda_{\rm z.m.}^{\dot\alpha}\sim \bar\tau^{\dot\alpha}$. More precisely, $h^0_Q(D)$ counts the number of holomorphic sections of $K_{\X}|_{\rm E3}$. Requiring $h^0_Q(D)=0$ implies that  $K_{\X}|_{\rm E3}$ must be a negative line-bundle on the E3-brane.

\bigskip

Finally, it can be useful to consider the weak coupling orientifold limit of the above results, in which the axion-dilaton can be approximated as constant and each O7-plane, covered by four D7-branes, generate a monodromy (\ref{O7D7monodromy}). This monodromy acts on the world-volume fermions $\tilde\lambda,\tilde\psi, \tilde\rho$ by reversing their sign, while it leaves  $\lambda,\psi,\rho$ invariant. By uplifting $\X$ to the double cover Calabi-Yau $\tilde\X$, one can see that the $\lambda,\psi,\rho$ are {\em even} under the orientifold involution, while $\tilde\lambda,\tilde\psi, \tilde\rho$ are {\em odd}. Then the fermionic zero modes are even/odd harmonic representatives of (complex conjugated of the) even/odd cohomology groups $H^{0,i}_\pm (\tilde D)$, where $\tilde D$ is the double cover of $D$:
\be\label{O7spectrum}
\begin{array}{c|c} \text{\lhdot\ zero modes} &\text{cohomology group}\\ \hline
\lambda^\alpha_{\rm z.m.}&  H_+^{0,0}(\tilde D)\\
\psi^\alpha_{\rm z.m.} & H_+^{0,1}(\tilde D)\\
\rho^\alpha_{\rm z.m.} & H_+^{2,0}(\tilde D)
\end{array}\qquad\begin{array}{c|c} \text{\rhdot\ zero modes} &\text{cohomology group}\\ \hline
\tilde\lambda^{\dot\alpha}_{\rm z.m.}&  H_-^{0,0}(\tilde D)\\
\tilde\psi^{\dot\alpha}_{\rm z.m.} &  H_-^{1,0}(\tilde D)\\
\tilde\rho^{\dot\alpha}_{\rm z.m.} &  H_-^{0,2}(\tilde D)\end{array}
\ee
This spectrum is  counted by
\be
h^{i}_\pm(D):=\dim H^{0,i}_\pm(D)
\ee
 and agrees with the results obtained in the literature based on specific orientifolds models. In particular, in the case in which  $h^1_\pm(D)=h^2_\pm(D)=0$ (that in particular implies that the divisor is rigid), the non-universal zero modes $\tilde\lambda_{\text{z.m.}}^{\dot\alpha}\sim \bar\tau^{\dot\alpha}$ are absent only if $h^{0,0}_-=0$, that is possible only if the E3-brane coincides with its orientifold image, \ie\  only if it is a so-called O(1) E3-brane \cite{Bianchi:2007fx, Argurio:2007qk,Argurio:2007vqa}.



\subsection{M5-brane index from IIB perspective}
\label{sec:index}

In \cite{witten96}, Witten provided a necessary condition for an E3-brane to contribute to the superpotential in terms of the arithmetic genus
of the divisor wrapped  by the dual M5-brane instanton. Here we would like to revisit Witten's argument from the IIB perspective. Of course, the two approaches must be equivalent, as we will discuss in the the following section.

Witten's argument is based on the observation that the M5-brane configuration is symmetric under a U(1) rotation along the normal bundle to the divisor wrapped by the M5-brane. In the IIB picture, one has to consider a rotation along the two directions transverse to the divisor $D\subset \X$.  In particular, one has to consider the action of this rotation on the  world-volume fermionic fields.

Let us denote with $({\ul 5},\ul 6)$ the two (flat) directions normal to the E3-brane. Rotations in these two directions can be seen as R-symmetry tranformations. We denote this symmetry by U(1)$_R$. The ten-dimensional spinorial action of the U(1)$_R$ rotation is generated by the matrix $R=\frac12\bbone\otimes \gamma_{\ul{56}}$, where we have used the gamma-matrices decomposition (\ref{gammadec}). By using (\ref{chiralm}) and (\ref{E3chiral}), one can alternatively write $R=-\frac12\bbone\otimes (\gamma_{\rm E3}\gamma_7)$. This generator acts in the same way on both components $\theta_1$ and $\theta_2$ of (\ref{GSfermion}). By looking at the expansion (\ref{Dfermsplit}) and using (\ref{E3susy}), it is easy to see that the world-volume fields transform with charges
\be\label{Rcharges}
\begin{array}{c|c} \text{\lhdot\ fermions} &\text{U(1)$_R$ charge}\\ \hline
\lambda^\alpha&  -1/2\\
\psi^\alpha & +1/2\\
\rho^\alpha & -1/2
\end{array}\quad \quad \begin{array}{c|c} \text{\rhdot\ fermions} &\text{U(1)$_R$ charge}\\ \hline
\tilde\lambda^{\dot\alpha}&  +1/2\\
\tilde\psi^{\dot\alpha} & -1/2\\
\tilde\rho^{\dot\alpha} & +1/2\end{array}
\ee

Taking into account the two-component spinorial structure of the world-volume zero-modes,  the path integral integration measure produces a violation of the U(1)$_R$ symmetry given by
\bea\label{difftwistedindex}
\chi_{\rm E3}&=& (h^{0}(D)- h^{1}(D)+ h^{2}(D)) - \big[h^{0}_Q(D)- h^{1}_Q(D)+h^{2}_Q(D)\big]\cr
&\equiv& \chi(D,\calo_D)-\chi(D,\call^{-1}_Q)
\eea
of the   U(1)$_R$ symmetry. As already written in (\ref{difftwistedindex}), $\chi_{\rm E3}$ can be seen as the difference of  the holomorphic Euler characteristics
\bea \label{diffindex}
\chi(D,\calo_D)&:=& h^{0}(D)- h^{1}(D)+ h^{2}(D)\cr
\chi(D,\call^{-1}_Q)&:= &h^{0}_Q(D)- h^{1}_Q(D)+h^{2}_Q(D)
\eea

The U(1)$_R$ symmetry should not be anomalous, that implies that the U(1)$_R$ violation generated by the fermion zero-modes must be compensated by a shift $T_{\rm E3}\rightarrow T_{\rm E3}-\ii\chi_{\rm E3}$ in the term $e^{-T_{\rm E3}}$ appearing in the non-perturbative contribution. In particular, a superpotential is generated only if in the complete path-integral  there are exactly two zero modes, that implies that a necessary condition  for having  a non-trivial superpotential is that
\be
\chi_{\rm E3}\equiv \chi(D, \calo_D)-\chi(D,\call^{-1}_Q)=1
\ee

In order to  evaluate $\chi_{\rm E3}$, it can be useful to express it in terms of characteristic classes by using the Hirzebruch-Riemann-Roch index theorem:
\be\label{index}
\chi(D,\calo_D)= \int_{D} {\rm Td}(D)\, ,\quad \chi(D,  \mathcal{L}_Q^{-1} ) = \int_{D} {\rm ch}( \call^{-1}_Q) \wedge {\rm Td}(D)
\ee
By using the isomorphism $\call_Q\simeq K_{\X}^{-1}$ and the adjunction formulas, the identities (\ref{index}) can be useful for  separately  computing $\chi(D,\calo_D)$ and $\chi(D,\call^{-1}_Q)$. For instance, the combination that appears in $\chi_{\rm E3}$ simplifies to
\begin{equation} \label{equivariantindex}
\chi_{\rm E3}= -\tfrac{1}{2}\, \int_{\X} c_1(\call_Q) \wedge  [D]\wedge [D]
\end{equation}
where $ [D]$ is the Poincar\'e dual to $D$ in $\X$.

By recalling the topology of the total divisor wrapped by the background 7-branes, it is immediate to see from  (\ref{equivariantindex}) that
\be\label{chiint}
\chi_{\rm E3}=-\frac{1}{24}n
\ee where $n$ is the total number of intersections of the 7-branes with the divisor self-intersection $D\cap D$, with the appropriate multiplicities. In particular, the tadpole conditions ensure that $n$ is a multiple of $24$.

Finally, notice that in the weak-coupling orientifold limit $\chi_{\rm E3}$ reduces to the equivariant index known as the holomorphic Lefschetz number $L(\tilde D)$ discussed in \cite{andres}
 \be
\chi_{\rm E3} = L(D)= h^{0}_+(\tilde D)-\big[h^{0}_-(\tilde D)+h^{1}_+(\tilde D)\big]+\big[h^{1}_-(\tilde D)+h^{2}_+(\tilde D)\big]-h^{2}_-(\tilde D)
\ee
that is computed in the double cover Calabi-Yau $\tilde\X$, with $\tilde D$ being the double cover of the divisor $D$.

In the orientifold limit,  the formula (\ref{chiint}) for $\chi_{\rm E3}$ can be expressed just in terms of intersections with the O7-planes. Let us continue working on $X=\tilde X/\mathbb{Z}_2$. Taking into account that each O7-plane is the bound-state of two mutually nonlocal 7-branes, one can write
\be
n=n_{\rm D7}+2n_{\rm O7}
\ee
On the other hand, the tadpole condition restricted on $D\cap D$ gives the constraint
\be
n_{\rm D7}=4n_{\rm O7}
\ee
Hence, $n=6n_{\rm O7}$ and one can write the index $\chi_{\rm E3}$ just  as
\bea\label{O7int}
\chi_{\rm E3}&=&-\frac14\, n_{\rm O7} =-\frac{1}{4}\, \int_{\X}[ D_{\rm O7}] \wedge  [D]\wedge [D] \cr
&=& -\frac{1}{4}\, \int_{\tilde\X}[ D_{\rm O7}] \wedge  [\tilde D]\wedge [\tilde D]
\eea
where in the second line we have rewritten the  result in the double cover Calabi-Yau $\tilde\X$.

More directly, in the weak coupling description reviewed in section \ref{sec:Mtheory}, the total divisor $D_{\rm O7}$ wrapped by the O7-planes is defined by (\ref{eqO7}), where $h$ is a section of $\call^2_Q$. Hence $[D_{\rm O7}]=c_1(\call_Q^2)=2c_1(\call_Q)$ and then (\ref{equivariantindex})  immediately implies the first line of (\ref{O7int}).

\subsection{Comparison with the dual M-theory result}
\label{sec:M5index}

In the past sections we have learned that the twisted theory on an E3-brane allows us to define the zero-modes as elements of sheaf cohomologies of the form $H^i(D, \calo_D)$ and $H^i(D, \call^{-1}_Q)$. In this section, we will briefly establish the link between these objects and the zero-modes of the dual M5-brane considered by Witten in \cite{witten96}. Most of this analysis is based on the papers \cite{grassi1, grassi2, kollar}, see also \cite{sethiNP}.

Witten's analysis translates the fermionic zero-modes on the M5-instanton in terms of its worldvolume $(0,i)$-forms. Given an E3-brane on a divisor $D \subset \X$, its dual M5-brane is simply the restriction onto $D$ of the elliptic fibration over $\X$. Hence, if $\pi$ is the projection of the elliptic fiber $\pi: \Y \rightarrow \X$, then the dual M5 wraps a divisor of the Calabi-Yau four-fold $\Y$ is defined as
\be
\hat D \equiv \pi^{-1}(D)
\ee
Since the fibration is holomorphic, the cohomologies of $D$ and $\hat D$ are related by the so-called Leray spectral sequence (see \cite{hartshorne}, and \cite{gh} for some material), that eventually implies that the Euler characteristics be related as follows:
\be
\chi(\hat D, \calo_{\hat D}) = \chi_{E3} \equiv \chi(D, \calo_D) - \chi(D, K_{\X})
\ee
Therefore, since $K_{\X}=\call^{-1}_Q$, the new IIB index (\ref{difftwistedindex})
computes exactly the index introduced by Witten in \cite{wittenM5} from the M-theory vantage point.

In order to learn about the zero-modes individually, however, the results of Koll\'ar \cite{kollar} are crucial, because they show that the Leray spectral sequence simplifies drastically\footnote{{\it I.e.}\ it degenerates at `page' $E_2$.} and one actually has the following relations:
\begin{eqnarray}\label{cohosplit}
\text{dim }H^0(\hat D, \calo_{\hat D}) &=& \text{dim }H^0(D, \calo_D) \nonumber \\
\text{dim }H^1(\hat D, \calo_{\hat D}) &=& \text{dim }H^0(D, K_\X)+\text{dim }H^1(D, \calo_D) \\
\text{dim }H^2(\hat D, \calo_{\hat D}) &=& \text{dim }H^1(D, K_\X)+\text{dim }H^2(D, \calo_D) \nonumber \\
\text{dim }H^3(\hat D, \calo_{\hat D}) &=& \text{dim }H^2(D, K_\X) \nonumber
\end{eqnarray}
that originally appeared in \cite{grassi1, grassi2}. Clearly, the cohomology groups in the r.h.s.\ of (\ref{cohosplit}) coincide with the (complex conjugated) cohomologies  appearing in (\ref{sheafcoho}), that were found from the direct IIB zero mode spectrum.

\begin{figure}[h]
\begin{center}
\begin{tikzpicture}[scale=1, node distance = 2cm, auto, inner sep=.1mm, text width=1.5cm, text centered]
\node (02m) at (0,.25) {$h^{2}_Q({\rm E3})$} ;
\node (02p) at (0,-.3) {$h^{2}({\rm E3})$};
\node (01m) at (0,-1.25) {$h^{1}_Q({\rm E3})$} ;
\node (01p) at (0,-1.8) {$h^{1}({\rm E3})$};
\node (00m) at (0,-2.75) {$h^{0}_Q({\rm E3})$} ;
\node (00p) at (0,-3.3) {$h^{0}({\rm E3})$};
\node[text width=4cm] at (-3,.25){\blue{\small geom.\ mod.}};
\node[text width=4cm] at (-3,-.3){\blue{\small tw.\ geom.\ mod.}};
\node[text width=4cm] at (-3,-1.25) {\blue{\small tw.\ Wilson lines}};
\node[text width=4cm] at (-3,-1.8) {\blue{\small Wilson lines}};
\node[text width=4cm] at (-3,-2.75) {\blue{\small $\bar \tau^{\dot\alpha}$}};
\node[text width=4cm] at (-3,-3.3) {\blue{\small $\theta^{\alpha}$}};
\node (03) at (4,1.3) {$h^{3}({\rm M5})$};
\node (02) at (4,-.2) {$h^{2}({\rm M5})$};
\node (01) at (4,-1.7) {$h^{1}({\rm M5})$};
\node (00) at (4,-3.2) {$h^{0}({\rm M5})$};
\path [line] (00p) -- (00);
\path [line] (00m) -- (01);
\path [line] (01p) -- (01);
\path [line] (01m) -- (02);
\path [line] (02p) -- (02);
\path [line] (02m) -- (03);

\end{tikzpicture}
\end{center}
\caption{\small Schematic description of the reorganization of the E3 fermionic zero modes in terms of the corresponding zero modes on the dual M5-brane. Here $h^i({\rm E3})\equiv h^i(D)$,  $h^i_Q({\rm E3})\equiv h^i_Q(D) = \dim H^i(D, K_X)$ and $h^{i}({\rm M5})\equiv\dim H^i(\hat D, \calo_{\hat D})\equiv\dim H^{0,i}_{\delbar}(\hat D)$.} \label{E3-M5}
\end{figure}
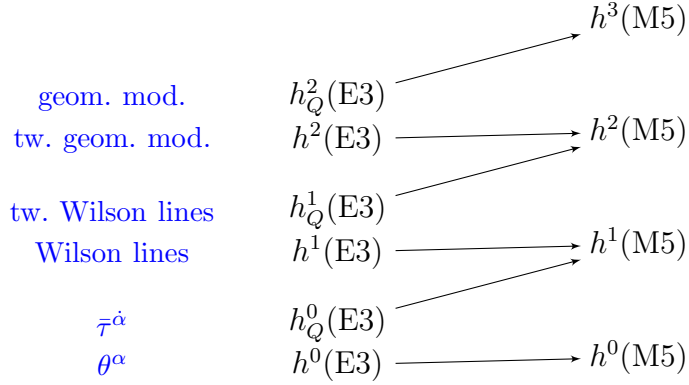

The relations (\ref{cohosplit}) were only understood as cohomological identities with no particular meaning.
The present treatment of the E3-instanton gives them physical significance and provides an interpretation of the M5 Hodge numbers in terms of more palatable quantities from the IIB perspective, that can be summarized by the diagram in Fig.\ \ref{E3-M5}. Therefore, given an instanton in a IIB setup, one does not necessarily need to construct an M/F-theory lift of it,
 in order to determine whether a non-perturbative superpotential will be generated.\footnote{It would be interesting to use the IIB/M-theory relation to explore the geometrization of non-perturbative effects proposed in \cite{tentofour,liam1,liam2,lucatolya} from the point of view of E3/M5-brane instantons.}

\subsection{$\chi_{\rm E3}$ as the perturbative Ext index}
\label{sec:ext}

In this section, we will show that the new index $\chi_{\rm E3}$, despite being defined in a non-perturbative context, can be understood in terms of objects that appear naturally in string theory, namely Ext groups.

A divisor $D\subset \X$ is associated to a holomorphic sheaf $\cale$ with support on $D$, defined by the short exact sequence
\be \label{skyscraper}
0 \rightarrow \calo_{\rm \X}(-D) \stackrel{\cdot P_D}\longrightarrow \calo_{\rm \X} \stackrel{|_{P_D=0}}\longrightarrow \cale_D \rightarrow 0
\ee
The first object corresponds to the negative bundle whose sections have degree opposite to $P_D$, where $P_D$ is the section of $ \calo_{\rm \X}(D)$ that vanishes on $D$. The middle map corresponds to multiplication by $P_D$, and the map into $\cale_D$ corresponds to setting $P_D=0$, i.e. restricting onto $D$. In mathematics, this is called a \emph{resolution} of $\cale_D$ by line-bundles (locally free sheaves).

One can now compute the groups ${\rm Ext}^i(\cale_D,\cale_D)$ -- see for instance \cite{sharperev,aspinwall} for an introduction to Ext-groups  and to their applications to D-brane physics. Using the so called local-to-global spectral sequence one gets
\begin{eqnarray} \label{extgroups}
\dim {\rm Ext}^0(\cale_D, \cale_D) &=& \dim H^0(D, \calo_D) \nonumber \\
\dim {\rm Ext}^1(\cale_D, \cale_D) &=& \dim H^1(D, \calo_D)+ \dim H^0(D, N_{D})\cr
&=&\dim H^1(D, \calo_D)+ \dim H^2(D, K_{\X})
\\
\dim {\rm Ext}^2(\cale_D, \cale_D) &=& \dim H^2(D, \calo_D)+ \dim H^1(D, N_{D})\nonumber \\
&=&\dim H^0(D, K_{\X})+ \dim H^1(D, K_{\X})\nonumber\\
\dim {\rm Ext}^3(\cale_D, \cale_D) &=& \dim H^2(D, N_{D})=\dim H^0(D, K_{\X})\nonumber
\end{eqnarray}
where $N_D$ is the holomorphic normal bundle to $D$.

By recalling that $K_{\X}\simeq \call^{-1}_Q$, one can see that the the ${\rm Ext}^i(\cale_D,\cale_D)$ classes coincide with the E3 fermionic zero-mode spectrum  (\ref{sheafcoho}). Furthermore, it is easy to see that $\chi_{\rm E3}$ reproduces the Ext-index\footnote{The Hirzebruch-Riemann-Roch theorem for Ext-groups reads $\int_{\X} {\rm ch}(\cale)\, {\rm ch}(\calf)^* \,{\rm Td}(\X)$, that can be used to further check the equivalence with $\chi_{\rm E3}$.}
\be\label{E3ext}
\chi_{\rm E3}\equiv {\rm Ind}_{\rm Ext}(\cale_D, \cale_D):=\sum_i (-)^i \dim {\rm Ext}^i(\cale_D, \cale_D)
\ee

In general, in the perturbative framework in which there are no SL$(2,\mathbb{Z})$-monodromies and $\X$  is a Calabi-Yau, given two holomorphic sheaves $\cale$ and $\calf$ on $\X$ associated to two D-branes,  the Ext groups   count the perturbative spectrum of open strings stretching between the two D-branes \cite{sharpe}. However, if $\X$ is a Calabi-Yau then  Serre-duality (\ie\ CPT conjugation) implies that  $\dim{\rm Ext}^i(\cale,\calf)=\dim {\rm Ext}^i(\calf,\calg)$ and then $ {\rm Ind}_{\rm Ext}(\cale_D, \cale_D)\equiv 0$.
In the present case, one gets in general a non-vanishing (\ref{E3ext}) because $K_{\X}\simeq \call_Q^{-1}$ is non-trivial, due to the fact that one is actually considering non-perturbative backgrounds.

In any case, it is somewhat surprising from a physical perspective that the Ext groups and the associated index, that are usually applied to {\em perturbative} D-brane physics,  capture the spectrum of the E3-brane in {\em non-perturbative} backgrounds. It is not completely clear to us whether  this just an accidental correspondence or  it  hides  a deeper motivation.

\subsection{Implementation in the working example}
\label{sec:example2}

Let us now put the results of this section into practice by using the example with $\X = \mathbb{P}^3$ that was introduced in \ref{sec:example}. We will first show them from the general IIB point of view on $\X$, then in Sen's limit from the double-cover point of view, and finally in terms of the dual M5-instanton.

Since this is a one-modulus three-fold , all divisors are Poincar\'e dual to some multiple of the hyperplane class $[H] \in H^2(\X, \mathbb{Z})$. Let use then choose as divisor wrapped by the E3-brane an hypersurface defined by
\be\label{exD}
D:\{P^{(n)}(z_1, \ldots, z_4)=0\}
\ee
 for any degree $n$ polynomial $P^{(n)}(z_1, \ldots, z_4)$. This fixes the cohomology class of its Poincar\'e dual to be
\be\label{cohoeq}
[D] \simeq n [H]
\ee

According to the discussion of section (\ref{sec:zeromodes}), in order to compute the spectrum of E3 fermionic zero-modes, one needs to compute $H^i(D, \calo_D)$ and $H^i(D, \call^{-1}_Q)$.

Let us first focus on the cohomology groups $H^i(D, \calo_D)$, that counts the zero modes associated to the\LH world-volume fields $\lambda,\psi,\rho$,  see (\ref{sheafcoho}).  In order to compute them, one can use the short exact sequence:
\be
0 \rightarrow \calo_{\mathbb{P}^3}(-n) \rightarrow \calo_{\mathbb{P}^3} \rightarrow \calo_D \rightarrow 0
\ee
where the first map is multiplication by the section of  $\calo_{\mathbb{P}^3}(n)$ that vanishes on $D$ and the second map is just  restriction.

This induces a long exact sequence of cohomologies. However, it is known that $H^i(\mathbb{P}^3, \calo_{\mathbb{P}^3}(k)) = 0$ for $i \neq 0, 3$ and any $k$. Hence the long sequence breaks into several short ones and  one eventually gets
\bea
\text{dim }H^0(D, \calo_D) &=& 1\cr
 \text{dim }H^1(D, \calo_D) &=& 0\cr
  \text{dim }H^2(D, \calo_D) &=& \text{dim }H^0(D, \calo_{\mathbb{P}^3}(n-4))
\eea
The dimension of $H^2(D, \calo_D) $ can be more explicitly computed by recalling that, in general, one can count the sections of a line bundle $\calo_{\mathbb{P}^3}(d)$ by counting monomials of degree $d$, that yields:
\be\label{formulaO(d)}
\text{dim }H^0(\mathbb{P}^3, \calo(d)) = \binom{ d+3}{ 3}
\ee

Hence, one obtains the following spectrum of\LH fermions on the E3-brane
\be\label{RHspectrum}
\begin{array}{c|c} \text{\lhdot\ fermions} &\text{\#\ zero modes}\\ \hline
\lambda^\alpha&  2\times 1\\
\psi^\alpha & 0\\
\rho^\alpha & 2\times \binom{ n-1 }{3}
\end{array}\begin{array}{r}\\ \\ \end{array}
\ee
As expected, for every $n$, there are the  two universal zero modes $\lambda^\alpha_{\rm z.m.}\sim \theta^\alpha$.
On the other hand, if $n\geq 4$, there are additional $\frac{2(n-1)!}{3!(n-4)!}$ zero-modes $\rho^\alpha_{\rm z.m.}$.

One can now turn to the\RH fermions $\tilde\lambda,\tilde\psi,\tilde\rho$, whose zero modes are counted by $H^i(D, \call_Q^{-1})\equiv H^i(D, \calo_{\mathbb{P}^3}(-4))$,  see (\ref{sheafcoho}). In order to compute their dimensions, one can follow the same strategy as above, using now the
short exact sequence
\be
0 \rightarrow \calo_{\mathbb{P}^3}(-n) \otimes K_{\mathbb{P}^3} \rightarrow K_{\mathbb{P}^3} \rightarrow K_{\mathbb{P}^3}|_D \rightarrow 0
\ee
The result is
\bea
\text{dim }H^0(D, K_{\mathbb{P}^3}) &=& 0\cr
 \text{dim }H^1(D, K_{\mathbb{P}^3}) &=& 0\cr
  \text{dim }H^2(D, K_{\mathbb{P}^3}) &=& \text{dim }H^0(\mathbb{P}^3, \calo_{\mathbb{P}^3}(n))-1
\eea
By using the general formula (\ref{formulaO(d)}), for the\RH fermionic spectrum
one then obtains
\be\label{LHspectrum}
\begin{array}{c|c} \text{\rhdot\ fermions} &\text{\#\ zero modes}\\ \hline
\tilde\lambda^{\dot\alpha}&  0\\
\tilde\psi^{\dot\alpha} & 0\\
\tilde\rho^{\dot\alpha} & 2\times \big[\binom{n+3 }{ 3}-1\big]
\end{array}\begin{array}{r}\\ \\ \end{array}
\ee
One can then see that the\RH sector contributes with $\frac{2(n+3)!}{n!3!}-2$ zero modes $\tilde\rho^{\dot\alpha}_{\rm z.m.}$.

In summary, in addition to the two universal zero modes $\lambda^\alpha_{\rm z.m.}\sim \theta^\alpha$ there is always
a certain number of zero modes $\tilde\rho^{\dot\alpha}_{\rm z.m.}$ and, for $n\geq 4$, additional zero modes $\rho^\alpha_{\rm z.m.}$.
The zero modes $\tilde\rho^{\dot\alpha}_{\rm z.m.}$ have a clear geometrical interpretation, as $ \text{dim}\,H^2(D, K_{\mathbb{P}^3})={\rm dim}\, H^0(D,N_D)$ by Serre duality and then they can be seen as the supersymmetric partner of the geometrical deformations of the divisor $D$.
Since $N_D\simeq \calo_{\mathbb{P}^3}(n)|_D$, they can be directly counted as follows:
\begin{eqnarray}\label{evendef}
\text{dim }H^2(D,K_{\mathbb{P}^3}) &=& \text{dim }H^0(\calo_{\mathbb{P}^3}(n)) \cr
&=& \text{\# monomials of degree n in } z_1, \ldots z_4 \text{ minus one}\nonumber\\
&=& \binom{n+3 }{ 3}-1
\end{eqnarray}
where the minus one corresponds subtracting the defining polynomial $P^{(n)}$, since we only count deformations.
On the other hand, $\rho^\alpha_{\rm z.m.}$ are associated to some kind of twisted geometrical deformations, whose interpretation will become more transparent in the weak coupling orientifold limit.

Now, the  index $\chi_{\rm E3}$ can be immediately computed from (\ref{difftwistedindex}) to be
\be\label{exE3index}
\chi_{\rm E3} = -2\,n^2
\ee
that is always different from $1$. Alternatively, one can easily get the same result  from  \eqref{equivariantindex}, by using (\ref{cohoeq}), together with the identity $c_1(\call_Q)=-c_1(K_{\mathbb{P}^3})=4[H]$ and the fact that $\int_{\mathbb{P}^3}[H]\wedge [H]\wedge [H]=1$.

\bigskip

Let us now revisit these results from the orientifold point of view, as was done in \cite{andres}. The divisor $D \subset \mathbb{P}^3$ has a double-cover divisor $\tilde D $ in the double-cover Calabi-Yau three-fold $\subset \tilde \X=\mathbb{P}^4_{1 1 1 1 4}[8]$ discussed in section \ref{sec:example}. Now, the canonical bundle of the three-fold  is trivial, but the relevant information is encoded in the $\mathbb{Z}_2$-equivariant structure of the orientifold involution $\xi \rightarrow -\xi$.

The condition (\ref{cohoeq}) translates into $[\tilde D] = n\,[H]$ in $\tilde \X$, where we continue to use the symbol $H$ to denote  the double cover of $H$, hoping that this will not cause confusion.
Its Poicar\'e dual  $[H]$ in $\tilde X$ can be thought of as the pullback of the hyperplane class $[H]$ in $\mathbb{P}^3$ under the $\mathbb{Z}_2$ projection and we use the notation $\calo_{\tilde \X}(n)\equiv \calo_{\tilde \X}(n\,H)$.  The most general effective divisor $\tilde D$ of this kind is given  by the intersection of (\ref{simpleCY}) and the vanishing locus
\be\label{coverdiv}
\tilde P^{(n)}(z_1,\ldots ,z_4,\xi)=0
\ee
of a homogeneous degree $n$ polynomial $\tilde P^{(n)}(z_1,\ldots ,z_4,\xi)$, where $\xi$ itself has degree four, that is furthermore even in
$\xi$ in order to respect the orientifold involution $\xi\rightarrow -\xi$. However, we restrict to configurations that are irreducible double covers of single connected holomorphic hypersurfaces $D$ in $\X\equiv \mathbb{P}^3$, with (\ref{coverdiv}) just given by (\ref{exD}), \ie\ independent of $\xi$.  This means that the divisor $\tilde D$ is transversely invariant  under the orientifold action,  \ie\ is a so-called O(1) instanton.

In order to understand the fermionic zero-mode spectrum, one needs to compute the orientifold-even and -odd cohomology groups $H^{i}_\pm(\tilde D,\calo_{\tilde D})$.

First, considering only connected transversely invariant divisors, one finds
\be
\text{dim } H^{0,0}_+(\tilde D) = 1\,, \quad \text{dim } H^{0,0}_-(\tilde D) = 0
\ee
This matches  (\ref{RHspectrum}) and (\ref{LHspectrum}), cf.\ with (\ref{O7spectrum}).

Furthermore, it is useful to note  that the line bundle $\calo_{\tilde \X}(1)$ is very ample. Then, the Lefschetz hyperplane theorem guarantees that
\be
H^1(\tilde D, \calo_{\tilde D}) = 0\quad\Leftrightarrow \quad H_\pm^{0,1}(\tilde D) = 0
\ee
By inspecting (\ref{O7spectrum}), one can see that this is again in agreement with (\ref{RHspectrum}) and (\ref{LHspectrum}).

Finally, by Serre duality, one knows that
\be
\text{dim }H^{0,2}_{\pm}(\tilde D) = \text{dim }H^0_{\mp}(\tilde D, N_{\tilde D})
\ee
where $N_{\tilde D}$ is the normal bundle to $\tilde D$ in $\tilde X$. Note that, since the holomorphic $(3,0)$ form is orientifold-odd, Serre duality exchanges the two subgroups of the cohomology with opposite $\mathbb{Z}_2$ parity.
Therefore, $H^{0,2}_{-}(\tilde D)$ corresponds to divisor deformations that respect the orientifold involution, whereas $H^{0,2}_{+}(\tilde D)$ corresponds to those that are odd under it. In fact, one can directly count the even and odd sections. Orientifold \emph{odd} sections of the normal bundle correspond to deformations of the divisor's polynomial $\tilde P^{(n)}(z_1,\ldots ,z_4,\xi)$ that are linear in $\xi$. The {\em even} ones do not contain $\xi$. Notice that  higher powers of $\xi$ can be eliminated via the hypersurface equation (\ref{simpleCY}).

It is then easy to see that the counting of $H^{0,2}_{-}(\tilde D)$ is identical to the counting  in (\ref{evendef}) and gives the same result. This is of course what one was to reproduce: the number of zero modes $\tilde\rho_{\rm z.m.}^{\dot\alpha}$ in (\ref{LHspectrum}).

On the other hand, the odd sections, that are linear in $\xi$, are counted by
\begin{eqnarray}
\dim H^{0,2}_{+}(\tilde D) &=& \text{dim }H^0_{-}(\tilde D, N_{\tilde D}) \cr
&=& \text{\# monomials of degree $n-4$ in } z_1, \ldots z_4 \nonumber\\
&=& \binom{n-4+3 }{ 3}
\end{eqnarray}
This  indeed coincides with the dimension of $H^2(D, \calo_D) $ and gives a clearer interpretation of the zero modes $\rho^\alpha_{\rm z.m.}$ found in (\ref{RHspectrum}).

\vskip 3mm

Finally, according to the discussion in section \ref{sec:M5index} one can easily  lift all this information to the dual M5-brane. By using (\ref{cohosplit}),  one can relate the cohomologies of $D \subset \X$ to the cohomologies of the M5-instanton divisor $\hat D \equiv \pi^{-1}(D) \subset \Y$, where $\pi: \Y \rightarrow \X$ is the projection  from the elliptically fibered Calabi-Yau four-fold to its base $\X$, as follows:
\begin{eqnarray}
\text{dim }H^0(\hat D, \calo_{\hat D}) &=& \text{dim }H^0(D, \calo_D) =1 \nonumber \\
\text{dim }H^1(\hat D, \calo_{\hat D}) &=& \text{dim }H^0(D, K_\X)+\text{dim }H^1(D, \calo_D) = 0 \\
\text{dim }H^2(\hat D, \calo_{\hat D}) &=& \text{dim }H^1(D, K_\X)+\text{dim }H^2(D, \calo_D) = \binom{n-1 }{ 3} \nonumber \\
\text{dim }H^3(\hat D, \calo_{\hat D}) &=& \text{dim }H^2(D, K_\X) = \binom{n+3 }{ 3}-1\nonumber
\end{eqnarray}
The relevant index for the M5, \ie\  the holomorphic characteristic $\chi(\hat D, \calo)$ \cite{witten96}, does obviously coincide with (\ref{exE3index}).


\section{Magnetized  E3-instantons}
\label{sec:fluxE3}

From now on, we would like to relax the simplifying assumption (\ref{fluxless}) and study  how a non-trivial world-volume flux
\be\label{fluxfull}
\calf:= 2\pi\alpha^\prime F_{\rm E3}-\iota^*B_{\it 2}\neq 0
\ee
can change the story. The supersymmetry of the E3-brane imposes that it still wrap an effective divisor $D$ and that the world-volume flux $\calf$ satisfy the anti-self-duality condition
\be\label{ASDF}
\calf=-*\calf
\ee
that can be alternatively expressed by the pair of conditions
\bea
\calf^{0,2}&=&0 \qquad (\text{holomorphy}) \cr
J\wedge \calf&=&0\qquad (\text{primitivity})
\eea
 This conditions can be derived from standard arguments, paying particular attention to the analytic
continuation necessary to extend the fermions to Euclidean signature, as in section \ref{sec:fluxlessferm}.\footnote{See for instance appendix D of \cite{tentofour} for a detailed general discussion, that includes the setting considered in this paper as a particular subcase.}

\subsection{World-volume fluxes and duality}
\label{sec:fluxduality}

As it is clear form (\ref{fluxfull}), the world-volume flux $\calf$ receives two kinds of contributions. One comes from the pull-back of the bulk NS-NS $B_{\it 2}$ field, and the other from the world-volume flux $F_{\rm E3}$. Both transform non-trivially under the SL(2,$\mathbb{Z}$)-duality transformation.

Let us first recall that $B_{\it 2}$ pairs up  with the R-R two form $C_{\it 2}$ to transform as a doublet under SL(2,$\mathbb{Z}$)-duality
\be \left(\begin{array}{c} C_{\it 2} \\
B_{\it 2}\end{array}\right)\rightarrow \left(\begin{array}{cc} a & b \\ c &
d\end{array}\right)\left(\begin{array}{c} C_{\it 2} \\ B_{\it 2}\end{array}\right)
\ee
As we are going to discuss, this transformation is consistent with the transformation of the world-volume flux $F_{\rm E3}$.

Let us then turn to the purely world-volume part $F_{\rm E3}$. First introduce the dual field-strength $F^D_{\rm E3}$ defined as
\be\label{dualfield}
F^D_{\rm E3}=-2\pi
\ii\,\frac{\delta S_{\rm E3}}{\delta F_{\rm E3}}
\ee
In  this section, as already stated, we  work in Euclidean signature. Thus most equations should  be considered as the analytical continuation of equations in Minkwskian signature, wherein their meaning would be more transparent.
In (\ref{dualfield}) the functional derivative is defined by $\delta S_{\rm E3}=\int_D \frac{\delta
S_{\rm E3}}{\delta F_{\rm E3}}\wedge \delta F_{\rm E3}$
and $S_{\rm E3}$ is the (Wick rotatated) general off-shell action
\be\label{offshellE3}
S_{\rm
E3}=\frac{1}{(2\pi)^3 \alpha'^2}\int_D \d^4\sigma
\sqrt{\det(h+e^{-\phi/2}\calf)}\, +\frac{\ii}{(2\pi)^3 \alpha'^2}\int_D \Big( C_{\it 4}+C_{\it 2}\wedge \calf+\frac12C_{\it 0}\calf\wedge \calf\Big)
 \ee

This action, and actually its full supersymmetric extension, is invariant under SL(2,$\mathbb{Z}$)-duality, under which
 $F^D_{\rm E3}$ and $F_{\rm E3}$ transform as a doublet -- see for instance the explicit discussion in \cite{kimura}.
 Hence, it is useful to introduce
 \be
 \calf^D:=2\pi\alpha^\prime F^D_{\rm E3}-\iota^*C_{\it 2}
 \ee
that  pairs with $\calf$ into a doublet under  SL(2,$\mathbb{Z}$)-duality:
\be \left(\begin{array}{c} \calf^D \\
\calf\end{array}\right)\rightarrow \left(\begin{array}{cc} a & b \\ c &
d\end{array}\right)\left(\begin{array}{c} \calf^D \\ \calf\end{array}\right)
\ee

It is then convenient to consider the combinations
\bea
\calf^-&:=&\ii e^{\phi}(\bar\tau \calf-\calf^D)\cr
\calf^+&:=&-\ii e^{\phi}(\tau \calf-\calf^D)
\eea
that transform as sections of $\call_Q$ and   $\bar \call_Q$ respectively under SL(2,$\mathbb{Z}$)-duality.

Now, the key observation is that in the case we are interested in, namely E3-branes wrapping a divisor $D$ with anti-self-dual flux (\ref{ASDF}), one has
\be\label{chargeflux}
\calf^-=2\calf\, ,\qquad \calf^+=0    \qquad\qquad\text{(for $*\calf=-\calf$)}
\ee
A quick check of this is obtained by using the fact that the supersymmetric E3-brane is calibrated, in generalized sense of \cite{koerber}. For the purpose of evaluating the on-shell value of $F^D_{\rm E3}$ from (\ref{dualfield}), this allows to use the simplified action obtained by substituting the DBI term in (\ref{offshellE3}) with the value provided by the generalized calibration $\Re(e^{\ii J})$, that is
\be\label{simplyE3}
S_{\rm E3}=\frac{1}{(2\pi)^3 \alpha'^2}\int_D\Big[\frac12 J\wedge J +\ii\Big(\frac12\,\tau\, \calf\wedge \calf + C_{\it 4}+C_{\it 2}\wedge \calf\Big)\Big]+\ldots
\ee
where the $\ldots$ on the right are contributions at least quadratic in terms that vanish on the supersymmetric configurations \cite{lucasup} and then they do not effectively contribute to the evaluation of the on-shell $F^D_{\rm E3}$. 

From (\ref{simplyE3}) and (\ref{dualfield}) one readily gets $2\pi\alpha^\prime F^D_{\rm E3}=\tau\calf+C_{\it 2}$, that indeed implies (\ref{chargeflux}).
Equations (\ref{chargeflux}) have the consequence that, in the case we are interested in, one should consider $\calf$ as taking values in $\call_Q$, that is
\be
e^{-\phi/2}\calf\in \Lambda^{1,1}\otimes L_Q
\ee
Furthermore, since $\delbar\calf=0$, then  $\delbar_Q (e^{-\phi/2}\calf)=0$ and $\calf$ identifies the following cohomology classes
\be\label{cohoF}
[e^{-\phi/2}\calf]\in H^{1,1}_{\delbar } (D,L_Q)\quad  \Leftrightarrow\quad [\calf]\in   H^{1}(D,T^*_{D}\otimes  \call_Q)
\ee
where $T^*_{D}$ is the holomorphic cotangent bundle on $D$.

At this point one has to face a little subtlety. Take the Minkowskian Bianchi identity $\d\calf=0$ (if $H_{\it 3}=0$) and assume that it is valid in the Euclidean case we are interested in. Being $\calf$ purely $(1,1)$, this naturally splits into $\delbar\calf=0$ and $\del\calf=0$. Now, as we saw above,  $\delbar\calf=0$ is SL$(2,\mathbb{Z})$-covariant, being  equivalent to $\delbar_Q (e^{-\phi/2}\calf)=0$. On the other hand, $\del\calf=0$  is {\em not} SL$(2,\mathbb{Z})$-covariant. This effect is accompanied by a failure of the expected field equation $\d \calf^D=\d(\tau\calf)=0$. Again, this can be split into two parts: $\delbar \calf^D=0$ and $\del\calf^D=0$. Since $\calf^D=\tau\calf$, the first is indeed automatically
implied by the Bianchi identity $\delbar\calf=0$ and then it is SL$(2,\mathbb{Z})$-covariant by itself. On the other hand,  the second is {\em incompatible} with the SL$(2,\mathbb{Z})$ non-covariant Bianchi Identity $\del\calf=0$, and furthermore it is {\em not}  SL$(2,\mathbb{Z})$-covariant by itself. Hence, by invoking the principle of SL$(2,\mathbb{Z})$-covariance, it is natural to consider the possibility that $\del\calf=0$ and $\del(\tau\calf)=0$ are both incorrect and should be substituted by a single SL$(2,\mathbb{Z})$-covariant  equation, that should reduce to $\del\calf=0$ when $\tau$ is constant. The most natural choice is
\be\label{covholBI}
\del_Q (e^{-\phi/2}\calf)=0
\ee
We may interpret this possibility as follows. In Minkowski signature  Bianchi identities and equations of motion are usually  put on different footings and only the second ones are influenced by the background axion-dilaton. On the other hand, switching to Euclidean signature, the supersymmetric anti-self-duality condition mixes Bianchi identities and equations of motion into an anti-holomorphic ($\delbar$) and holomorphic ($\del$) part. Now it would be the anti-holomorphic part that is insensitive to the non-trivial $\tau$, while the holomorphic one would be modified by it.\footnote{This unbalancing between holomorphic and anti-holomorphic sectors  generated by the non-trivial $\tau$, that is an ubiquitous ingredient in the present discussion,  has a simpler manifestation if one considers E($-1$)-instantons. Indeed, the action would simply be given by $\tau(z_{\rm E(-1)})$, and so it would automatically be extremized wrt the anti-holomorphic variables ($\bar z_{\rm E(-1)}$) but not wrt the holomorphic ones ($z_{\rm E(-1)}$). } In fact, we will never explicitly need (\ref{covholBI}), with the only exception of section \ref{sec:lift}, where one will be naturally led to consider it.

The present situation is reminiscent of the fate of YM instantons in the presence of non-trivial VEV's for charged scalars. Instanton solutions no longer satisfy the classical coupled field equations and they only extremize the Euclidean action if some `constraint' is added by including additional terms in the action, whence the name `constrained instantons'. Although one can feel uneasy by expanding around a configuration that is not a solution of the `naive' field equations, the success of supersymmetric instanton calculus\footnote{See \eg \cite{Bianchi:2007ft} for a recent review.}, that relies on elegant and sound localization techniques, should be taken as an encouraging analogy for the validity of our analysis. At any rate, whenever a perturbative orientifold limit with constant $\tau$ is possible, these subtleties seems to disappear. Furthermore, even for weakly coupled orientifold limits with non-constant $\tau$,  one could check the validity of the analysis by directly using open-string techniques.


\subsection{Fermions and fluxes: the simplest case of constant $\tau$}
\label{sec:fluxconstau}

We now come back to our main problem, the study of the effect of a non-trivial world-volume flux (\ref{fluxfull}) on the fermionic zero-modes
spectrum.  In this subsection we restrict to the simpler subcase in which the axion-dilaton is (or is approximated to be) constant and the internal metric is Ricci-flat. Clearly the weak-coupling orientifold backgrounds fit into this class.

What one needs to do is to repeat the steps of section \ref{sec:fluxlessferm}, so one can proceed quite speedily, {\em mutatis mutandis} for taking into account the non-trivial flux. First of all, the operator $\Gamma_{\rm E3}$ gets modified. Namely,  by using the gamma-matrix decomposition (\ref{gammadec}),
in (\ref{gammaE3})  one should take $\hat \gamma_{\rm E3}=\bbone\otimes \gamma_{\rm E3}(\calf)$, with
\be\label{fluxgamma}
\gamma_{\rm
E3}(\calf)=\frac{-\ii\,\epsilon^{a_1\ldots a_4}}{\sqrt{\det
(h+e^{-\phi/2}\calf)}}\Big(\frac{1}{4!}\gamma_{a_1\ldots
a_4}+\frac{1}{4}\,e^{-\phi/2}\calf_{a_1a_2}\gamma_{a_3
a_4}+\frac{1}{8}\,e^{-\phi}\calf_{a_1a_2}\calf_{a_3
a_4}\Big) \ee
We continue imposing $\kappa$-fixing condition in the form (\ref{genkappafix}) and the
supersymmetry condition still reads (\ref{E3chiral}).

The quadratic action for the $\kappa$-fixed Green-Schwarz fermions is now given by \cite{Ddirac}
\be\label{thetactionflux}
S_{\rm F}=\frac{2\pi \ii}{\ell^4_s}\int_{D}
\d^4\sigma\,\sqrt{\det M} \, \,\overline{\Theta}(\calm^{-1})^{ab}\Gamma_a \hat \nabla_b\Theta
\ee
where $M_{ab}:=h_{ab}+e^{-\phi/2}\calf_{ab}$ and $\calm_{ab}:=h_{ab}\bbone+e^{-\phi/2}\calf_{ab}\,\sigma_3$.

One also needs to decompose $\Theta$ in terms of new (topologically twisted) world-volume fields. Because of the modification of $\Gamma_{\rm E3}$,  the decomposition  (\ref{Dfermsplit}) does not fulfill the $\kappa$-fixing condition (\ref{genkappafix}) anymore and must be deformed into a new one. A new decomposition that we will find convenient is
\bea\label{Dfermsplitflux}
&&\left\{ \begin{array}{l} \theta_1^\R=\frac12\Big(\tilde\lambda\otimes \eta+\sqrt{\frac{\det h}{\det M}}M^a{}_b\tilde\psi_a\otimes \gamma^b\eta^*+\frac12\tilde\rho_{ab}\otimes \gamma^{ab}\eta \Big)
\\
\theta_2^\R=\frac{\ii}2\Big(\tilde\lambda\otimes \eta-\sqrt{\frac{\det h}{\det M}}(M^T)^a{}_b\tilde\psi_a\otimes \gamma^a\eta^*+\frac{1}2\tilde\rho_{ab}\otimes \gamma^{ab}\eta\Big)
\end{array}  \right. \cr
&& \left\{ \begin{array}{l}
\theta_1^\L=\frac12\Big(\lambda\otimes \eta^*+\sqrt{\frac{\det h}{\det M}}M^a{}_b\psi_a\otimes \gamma^b\eta+\frac12\rho_{ab}\otimes \gamma^{ab}\eta^*\Big)
 \\
\theta_2^\L=\frac{\ii}{2}\Big(\lambda\otimes \eta^*-\sqrt{\frac{\det h}{\det M}}(M^T)^a{}_b\psi_a\otimes \gamma^a\eta+\frac{1}2\rho_{ab}\otimes \gamma^{ab}\eta^*\Big)
\end{array}  \right.
\eea
The decomposition (\ref{Dfermsplitflux}) clearly reduces to (\ref{Dfermsplit}) in the limit $\calf\rightarrow 0$, in which $M\rightarrow h$.

By pugging (\ref{Dfermsplitflux}) into the action (\ref{thetactionflux}), after some manipulations, one gets
\bea\label{actionflux}
S_{\rm F}&=&\frac{4\pi\ii }{\ell_s^4}\int_{D} \big(\psi\wedge
*\del\lambda-\tilde\psi\wedge *\delbar\tilde\lambda-\rho\wedge*\delbar\psi+\tilde\rho\wedge*\del\tilde\psi\big)\cr
&&\quad\qquad+\frac{4\pi\ii }{\ell_s^4}\int_{D}\sqrt{\det h} \,\big(\rho\cdot \cals_\calf\cdot \rho -\tilde\rho\cdot\tilde\cals_\calf\cdot \tilde\rho\big)
\eea
We have introduced the short-handed notation
\be
\rho\cdot \cals_\calf\cdot \rho:= \frac14 \rho^{ab}(\cals_{\calf})_{ab}{}^{cd}\rho_{cd}\, ,\quad
 \tilde\rho\cdot \tilde\cals_\calf\cdot \tilde\rho:= \frac14 \tilde\rho^{ab}(\tilde\cals_{\calf})_{ab}{}^{cd}\tilde\rho_{cd}
 \ee
  where
$\cals_\calf$ and $\tilde\cals_{\calf}$ are tensors that are anti-symmetric and purely (anti-)holomorphic in $(ab)$ and $(cd)$, explicitly given in (anti-)holomorphic indices by
\bea
(\cals_\calf)_{\bar\imath\bar\jmath}{}^{uv}&=&-e^{-\phi}\calk^m{}_{\bar t[\bar\imath}\calf_{\bar\jmath]}{}^{\bar t}
\bar\Omega_{m}{}^{uv}\cr
(\tilde\cals_\calf)_{ij}{}^{\bar u\bar v}&=&-e^{-\phi}\calk^m{}_{t[i}\calf_{j]}{}^{t}
\Omega_{m}{}^{\bar u\bar v}\
\eea

The first line in (\ref{actionflux}) is just  (\ref{twaction2}) in the case of constant axion-dilaton, $Q_{\it 1}=0$. Hence, one can see that the new decomposition (\ref{Dfermsplitflux}) allows us to get the same kinetic terms for the topologically twisted world-volume fields. The possibility to reach this result
is not a priory obvious at all.

The second line of (\ref{actionflux}) provides a new mass-like term for $\rho$ and $\tilde\rho$, induced by the world-volume flux through the appearance of the extrinsic curvature $\calk^m{}_{ab}$. The definition and some useful properties of the extrinsic curvature are reviewed in appendix \ref{app:extrcurv}.

In the case of an orientifold background, the action of the E3-brane wrapping the double-cover divisor $\tilde D$ in the double cover Calabi-Yau $\tilde X$ take the very same form as (\ref{actionflux}). Both $\Omega$ and $\calf$ are odd under the orientifold involution and then $S_\calf$ and $\tilde S_\calf$ stay invariant. Furthermore, since the monodromy (\ref{O7D7monodromy})  acts  on $\Theta$ as $-\ii\sigma_2$ and on $\eta$ as $\eta\rightarrow \ii\eta$,  it is easy to see from (\ref{Dfermsplitflux}) that $\lambda$, $\psi$ and $\rho$ are even under the O7-involution,  while $\tilde\lambda$, $\tilde\psi$ and $\tilde\rho$ are odd. Hence the action (\ref{actionflux}) is manifestly invariant under the orientifold involution, as required.

We stress that the above transformation rules under the orientifold involution for the topologically twisted  world-volume fermions are the same as those found in the flux-less case discussed in section \ref{sec:zeromodes}. This is probably one of the reasons why our choice (\ref{Dfermsplitflux}) leads to the particularly simple action (\ref{actionflux}). Crucially, as we will see in the next subsection, this observation continue to hold for the complete SL(2,$\mathbb{Z}$)-duality group.

\subsection{Fermions and fluxes with generic $\tau$}

We can now address the generalization of the above results to the case in which $\tau$ has a more general non-constant (and holomorphic) profile. In principle, one could repeat the steps of the previous subsection, starting back again from the general action in \cite{Ddirac} and simply allowing for a non-constant $\tau$. Although straightforward, this procedure turns out to be technically quite intricate.
Then, here we follow an alternative strategy, that allows to obtain the desired result without much effort.

The important ingredient is that the action should be invariant under SL(2,$\mathbb{Z}$)  duality. This has been demonstrated in full generality in \cite{kimura} and in this section we would like to explore the consequences of this general result for the specific problem at hand.
In the flux-less case of section \ref{sec:fluxlessE3} we found that the topologically twisted world-volume fermions transform nicely under the SL(2,$\mathbb{Z}$)  duality and in section \ref{sec:fluxconstau} we have already seen that, for constant $\tau$, this property is preserved for the $\mathbb{Z}_2$ subgroup corresponding to the orientifold involution. What we would like to argue here is that, even for non-constant $\tau$, the decomposition (\ref{Dfermsplitflux}) provides topologically twisted world-volume fermions that continue to transform  under the SL(2,$\mathbb{Z}$)  duality in exactly the same simple way that was found in the flux-less case.

The understanding of the transformation properties of the topologically twisted world-volume fermions is not as straightforward as in the
fluxless case for the following reason.  As already mentioned, in general the combinations $\theta_1\pm \ii\theta_2$ of the Majorana-Weyl fermions appearing in (\ref{GSfermion}) transform with U(1)$_Q$ charge $\pm1/2$ under SL(2,$\mathbb{Z}$)  duality. If one tries to naively apply this transformation to (\ref{Dfermsplitflux}), taking into account that $\eta$ and $\calf$ have U(1)$_Q$ charge $+1/2$ and $+1$, one seems to encounters an horrible non-linear propagation of this U(1)$_Q$ action on the topologically twisted world-volume fermions, that originates exactly in the (non-linear) presence of the non-vanishing $\calf$ in  (\ref{Dfermsplitflux}).
However, one has to be careful since, in presence of world-volume fluxes, this U(1)$_Q$  action generically breaks the $\kappa$-fixing condition
(\ref{genkappafix}). Let us try to be very explicit on this point.

In bi-spinor formalism, the U(1)$_Q$ action is given by
\be\label{2compS}
\Theta\rightarrow \Theta'=e^{-\frac{\ii}2\alpha\,\sigma_2 }\Theta
\ee
where $\alpha=\arg(c\tau+d)$.
On the other hand, the infinitesimal $\kappa$-symmetry gauge transformation around a bosonic background can be written as
\be\label{2compkappa}
\delta_\kappa\Theta=P^+_\kappa(\calf)\kappa
\ee
for an arbitrary spinor $\kappa$, where use of the projectors has been made
\be
P^\pm_\kappa(\calf)=\frac12\big(\bbone\pm \Gamma_{\rm E3}\big)
\ee
that, as explicitly indicated, depend on $\calf$.
In this paper we have chosen to gauge-fix (\ref{2compkappa}) by imposing that
\be\label{kfix2}
P^+_\kappa(\calf)\Theta=0
\ee
Now, it should be evident that the SL(2,$\mathbb{Z}$)  action (\ref{2compS}) generically breaks (\ref{kfix2}),
that must then be re-established by acting with a compensating $\kappa$-symmetry transformation.

In order to find the $\kappa$-fixing preserving transformation, it is convenient to work with an infinitesimal SL(2,$\mathbb{R}$)-trasformation of the form (\ref{2compS}), that should still be preserved at the supergravity level. Hence
\be
\delta_S\Theta=-\frac{\ii}2\,\alpha\, \sigma_2\Theta
\ee
Now, the deformed bispinor $\Theta^\prime\simeq\Theta+\delta_S\Theta$ should be corrected by a $\kappa$-symmetry transformation  in such a way that it preserves the $\kappa$-fixing condition $P^+_\kappa(\calf^\prime)\Theta^\prime=0$, where $e^{-\phi^\prime/2}\calf^\prime=e^{\ii\alpha}e^{-\phi/2}\calf\simeq (1+\ii\alpha)e^{-\phi/2}\calf$. Hence, by defining $\delta_S P^-_\kappa(\calf)=P^-_\kappa(\calf^\prime)-P^-_\kappa(\calf)$, the $\kappa$-fixing preserving first order deformation is given by
\be\label{modSdef}
\hat\delta_S\Theta=P^-_\kappa(\calf)\, \delta_S\Theta+\delta P^-_\kappa(\calf)\, \Theta
\ee

At this point, all one has to do is to plug the decomposition (\ref{Dfermsplitflux}) into (\ref{modSdef}), taking into account that $\eta$ transform with U(1)$_Q$ charge $+1/2$ and extract the resulting transformations for the world-volume fermions. After some tedious algebra, one eventually finds that all nasty terms drop out and one is left with a simple linear action of the topologically twisted world-volume fermions
\bea
&&\hat\delta_S\lambda=\hat\delta_S\psi=\hat\delta_S\rho=0\cr
&&\hat\delta_S\tilde\lambda=-\ii\alpha\tilde\lambda\, ,\quad \hat\delta_S\tilde\psi=\ii\alpha\tilde\psi\, ,\quad \hat\delta_S\tilde\rho=-\ii\alpha\tilde\rho
\eea
Hence, on the twisted world-volume fermions the apparently complicated action of the infinitesimal SL(2,$\mathbb{R}$) duality simplifies drastically and reduces to a simple linear action. In particular, this can be readily exponentiated into a simple finite  SL(2,$\mathbb{R}$), and hence  SL(2,$\mathbb{Z}$), action  of the form
\be\label{Saction}
\lambda,\psi,\rho\rightarrow \lambda,\psi,\rho \quad,\quad \tilde\lambda\rightarrow e^{-\ii\alpha}\tilde\lambda \quad,\quad \tilde\psi\rightarrow e^{\ii\alpha}\tilde\psi \quad,\quad \tilde\rho\rightarrow e^{-\ii\alpha}\tilde\rho
\ee
Hence, even in presence of fluxes, the topologically twisted world-volume fermions continue to transform under SL(2,$\mathbb{Z}$)-duality with the same U(1)$_Q$ charges as in the fluxless case. In particular, their properties are still summarized by (\ref{Qcharge}).
One can then see the naturalness of the decomposition (\ref{Dfermsplitflux}), even in the case of non-constant $\tau$.

Furthermore, notice that the tensors $\cals_\calf$ and $\tilde S_\calf$ transform nicely under SL(2,$\mathbb{Z}$)-transformation, namely with charge $0$ and $+2$. In particular, they can be seen as operators
\be
\cals_\calf: \Lambda^{2,0} \rightarrow \Lambda^{0,2}\, ,\quad \tilde\cals_\calf:  \Lambda^{0,2}\otimes L_Q^{-1}\rightarrow \Lambda^{2,0} \otimes L_Q
\ee

One can now make the following observation. Take $\tau$-constant. With the above rules, it is clear that the action  (\ref{actionflux}) is invariant under a global SL(2,$\mathbb{Z}$)-duality. This action must extend to the case of a more general F-theory background with non-constant $\tau$ in such a way as to be invariant under the SL(2,$\mathbb{Z}$)-transformations that characterize it. Furthermore this generalization must reduce to (\ref{twaction2}) in the case of constant flux. All these ingredients mashed together practically uniquely fix the complete effective
action to be the simple `covariantization' of (\ref{actionflux}) obtained by replacing the ordinary derivatives on U(1)$_Q$-charged fields with the U(1)$_Q$-covariant derivatives.

Hence, we conclude that, in presence of a non-trivial world-volume flux, the fermionic E3-brane action is given by
\bea\label{actionflux2}
S_{\rm F}&=&\frac{4\pi\ii }{\ell_s^4}\int_{D} \big(\psi\wedge
*\del\lambda-\tilde\psi\wedge *\delbar_Q\tilde\lambda-\rho\wedge*\delbar\psi+\tilde\rho\wedge*\del_Q\tilde\psi\big)\cr
&&\quad\qquad+\frac{4\pi\ii }{\ell_s^4}\int_{D}\sqrt{\det h} \,\big(\rho\cdot \cals_\calf\cdot \rho -\tilde\rho\cdot\tilde\cals_\calf\cdot \tilde\rho\big)
\eea
where $\del_Q:=\del -\ii qQ^{1,0}$, $\delbar_Q:=\delbar -\ii qQ^{0,1}$, with $q$ being the U(1)$_Q$-charge given in (\ref{Qcharge}).

\subsection{Flux-modified zero modes}

The action (\ref{twaction2}) leads to the following flux-modified fermionic equations
\bea\label{Feomflux}
 \del\lambda=0\quad &,& \quad \delbar_Q\tilde\lambda=0 \cr
  \delbar^\dagger\psi=0\quad &,& \quad \del_Q^{\dagger}\tilde\psi=0 \cr
\del^{\dagger}\rho=0\quad &,& \quad \delbar_Q^{\dagger}\tilde\rho=0\cr
 \delbar\psi=\cals_\calf\cdot\rho
\quad &,& \quad  \del_Q\tilde\psi=\tilde\cals_\calf\cdot\tilde\rho
 \eea
Clearly, comparing these equations with the flux-less ones (\ref{Feom}), one can see that the only effect of the flux is to
slightly mix $\psi$ and $\rho$, and analogously for $\tilde\psi$ and $\tilde\rho$. However, in order to count the zero modes associated to $\psi$ and $\tilde\psi$, one can consistently set $\rho=\tilde\rho=0$. So one obtains that $\psi_{\rm z.m.}$ and $\tilde\psi_{\rm z.m.}$ are still given by harmonic representatives of the cohomology groups $H^{0,1}_{ \delbar} (D)$ and $H^{1,0}_\del(D,L_Q)$ respectively. Nothing changes for $\lambda_{\rm z.m.}$ and $\tilde\lambda_{\rm z.m.}$ as well, that are still given by the harmonic representatives of $H^{0,0}_\del(D)$ and $H^{0,0}_{\delbar }(D,L^{-1}_Q)$.

The story changes for $\rho$ and $\tilde\rho$.
On the one hand, the third line of (\ref{Feomflux})
implies that $\rho_{\rm z.m.}$ and $\tilde\rho_{\rm z.m.}$ must still be a {\em harmonic} representatives of $H^{2,0}_\del(D)$ and $H^{0,2}_{\delbar}(D,L^{-1}_Q)$ respectively, as in the flux-less case. However, for  $\rho_{\rm z.m.}$ and $\tilde\rho_{\rm z.m.}$ to be true zero-modes, the last line in (\ref{Feomflux}) requires $\cals_\calf\cdot\rho_{\rm z.m.}$ to be $\delbar$-exact and  $\tilde\cals_\calf\cdot\tilde\rho_{\rm z.m.}$ must be $\del_Q$-exact.

One can rewrite this condition as follows. First,  restricting the action of $\cals_\calf$ and $\tilde S_\calf$ onto harmonic forms, one can see them as maps between cohomology classes
\bea
\cals_\calf&:& H^{2,0}_\del(D)\rightarrow H^{0,2}_{\delbar}(D)\cr
\tilde\cals_\calf &:& H^{0,2}_{\delbar}(D, L_Q^{-1})\rightarrow H^{2,0}_{\del}(D,L_Q)
\eea
Hence, the condition for harmonic $\rho_{\rm z.m.}$ and $\tilde\rho_{\rm z.m.}$ to be true zero-modes is that
\bea\label{fluxcohocond2}
[\cals_\calf\cdot\rho]=0&\in& H_{\delbar}^{0,2}(D)\cr
[\tilde\cals_\calf\cdot\tilde\rho]=0&\in& H_\del^{2,0}(D,L_Q)
\eea

Once (\ref{fluxcohocond2}) is guaranteed, the last equation in (\ref{Feomflux}) can be integrated by using the standard
Hodge decomposition
\bea\label{hodgedec}
\psi&=&\psi_{\rm z.m.}+\delbar^\dagger\Delta_{\delbar}^{-1}(\cals_\calf\cdot\rho_{\rm z.m.})\cr
\tilde\psi&=&\tilde\psi_{\rm z.m.}+\del^\dagger_Q\Delta^{-1}_{\del_Q}(\tilde\cals_\calf\cdot\tilde\rho_{\rm z.m.})
\eea
Here $\psi_{\rm z.m.}$ and $\tilde\psi_{\rm z.m.}$ are the harmonic forms introduced above, that count as independent zero modes. On the other hand, one can see that  if $\rho_{\rm z.m.}$ ($\tilde\rho_{\rm z.m.}$) is non-vanishing and satisfies the condition (\ref{fluxcohocond2}), then $\psi$ ($\tilde\psi$) acquires an additional corresponding non-vanishing but uniquely determined profile.

One could re-express the above results in terms of sheaf cohomologies. Namely, one can see $\cals_\calf$  and $\tilde S_\calf$
as maps
\bea
\cals_\calf&:&   H^2(D,\bar\calo)\rightarrow   H^{2}(D,\calo)\cr
\tilde\cals_\calf &:&   H^{2}(D, \call_Q^{-1})\rightarrow   H^{2}(D,\bar\call^{-1}_Q)
\eea

Then, in summary, one finds that the in presence of fluxes the fermionic zero modes are associated with
\bea\label{sheafcoho2}
\begin{array}{c|c} \text{\lhdot\ zero modes} &\text{vector space}\\ \hline
\lambda^\alpha_{\rm z.m.}& H^0(D,\bar\calo_D)\\
\psi^\alpha_{\rm z.m.} &  H^{1}(D,\calo_D)\\
\rho^\alpha_{\rm z.m.} &  \ker \cals_\calf \subset   H^{2}(D,\bar\calo_D)
\end{array}\qquad \begin{array}{c|c} \text{\rhdot\ zero modes} &\text{vector space}\\ \hline
\tilde\lambda^{\dot\alpha}_{\rm z.m.}&  H^0(D, \call_Q^{-1})\\
\tilde\psi^{\dot\alpha}_{\rm z.m.} &  H^{1}(D,\bar\call^{-1}_Q)\\
\tilde\rho^{\dot\alpha}_{\rm z.m.} & \ker \tilde\cals_\calf\subset   H^{2}(D, {\cal L}_Q^{-1})\end{array}
\eea

These results clearly show that a world-volume flux can potentiallly lift part of the zero-modes of $\rho$ and $\tilde\rho$.
According to the discussion of section \ref{sec:zeromodes}, these zero-modes are the ($Q$-twisted) supersymmetric partners of the bosonic zero modes describing the infinitesimal deformations of the divisor. In fact, the above flux-induced fermionic moduli lifting has a geometrical counterpart -- see for instance \cite{sharpe,sharperev,lucadef}. We will come back to this point in subsection \ref{sec:lift}

On the other hand, one can also see  very explicitly that world-volume fluxes {\em cannot} lift the $h^0(D)$, $h^1(D)$, $h^0_Q(D)$ and $h^1_Q(D)$ zero modes. This observation fits in well with the results scattered in the literature, mostly based on specific examples -- see for instance \cite{ralphreview} and references therein.

\subsection{Index for magnetized E3-branes}
\label{sec:fluxindex}

In section \ref{sec:index} we have provided a direct IIB derivation of the index in \cite{witten96}. The derivation is based on the U(1)$_R$ symmetry correponding to a rotation of the complex coordinate orthogonal to the divisor. Notice that, even in presence of fluxes, the  U(1)$_R$  charges of the topologically twisted world-volume fermions are still given by (\ref{Rcharges}).
Hence by looking at the quadratic fermionic action (\ref{actionflux2}) it is immediate to realize that, contrary to the naive expectation,
 the presence of world-volume flux explicitly breaks the U(1)$_R$ symmetry. The breaking has its origin in the appearance of the extrinsic curvature. However, even though the U(1)$_R$ symmetry acting only on the word-volume fermions is broken, the extrinsic curvature appears contracted with
three-form $\Omega$ or its complex conjugated. As it is evident from (\ref{omega}), $\Omega$ transforms with U(1)$_R$ charge $-1$.
Hence, from the world-volume perspective, $\Omega$ can be thought of as a spurion restoring the U(1)$_R$ that, regarded as a local Lorentz symmetry, should not be broken in the complete theory.

Such spurionic couplings give a mass to the $\rho$ and $\tilde\rho$ zero modes that lie in the kernel of the operators $\cals_\calf$ and $\tilde\cals_\calf$ defined above.  Hence, in the path-integral, the  integration of these massive would-be zero-modes  will pull-down such spurionic mass terms that, combined with the factor $e^{-T_{\rm E3}}$, will modify the anomaly argument.

Hence  by  defining, in addition to  (\ref{hodgens}),  the flux-modified Hodge numbers
\be
h^{2}(\calf ) := {\rm dim}\, [\ker \cals_\calf \subset   H^{2}(D,\bar\calo)]\, ,\quad
h^2_Q(\calf):= \dim [\ker \tilde\cals_\calf\subset   H^{2}(D, {\cal L}_Q^{-1})]
\ee
one is  led to consider the modified index
\bea\label{fluxtwistedindex}
\chi_{\rm E3}(\calf)&:=& h^{0}-(h^{0}_Q+h^{1})+[h^{1}_Q+h^{2}(\calf)]-h^{2}_Q(\calf)\cr
&\equiv& \chi^0_{\rm E3}-(\dim\Im \cals_\calf-\dim\Im \tilde \cals_\calf)
\eea
where we have defined $\chi^0_{\rm E3}: =\chi_{\rm E3}(\calf=0)$.

Clearly $\chi_{\rm E3}(\calf)$ counts the amount of surviving zero-modes weighted by the sign of their  U(1)$_R$ charge.
The natural generalization of Witten's criterion, necessary but by no means sufficient, is then given by
\be\label{fluxcrit}
\chi_{\rm E3}(\calf)=1
\ee

So far, we have only considered the quadratic fermionic action. Hence, it could appear that in principle there could be other  U(1)$_R$ violating higher order terms that could spoil (\ref{fluxcrit}). However, standard non-rinormalization arguments imply that, up to the overall factor $e^{-T_{\rm E3}}$, the resulting superpotential can be computed in the limit $T_{\rm E3}\rightarrow \infty$, in which the metric of the bulk three-fold  $X$ is infinitely scaled up. In this limit the higher order terms vanish and hence we do not expect them to modify the above arguments.
Under this reasonable assumption, (\ref{fluxcrit}) appears as the most natural necessary condition to be considered.

Finally, in section \ref{sec:ext} we have pointed out the identity  (\ref{E3ext}) between $\chi_{\rm E3}$ in the absence of fluxes and the Ext-index. It is not clear to us whether this correspondence can be in some way extended to the case in which world-volume fluxes are turned on. One of the reasons is that, in an F-theory background, the flux cannot be easily associated to standard perturbative data underlying the definition of Ext-groups.

It is conceivable that one might be able to calculate the index \eqref{fluxcrit} in the orientifold weak coupling limit via some generalization of the $\mathbb{Z}_2$-equivariant Hirzebruch-Riemann-Roch index theorems exploited in \cite{Brunner:2003zm}, but for Ext-groups.


\subsection{On flux-induced zero-modes lifting}
\label{sec:lift}

We would like to understand better the mechanism that regulates the flux-induced moduli-lifting. As already mentioned, this mechanism has a geometric counterpart.

The infinitesimal deformations of the divisor are generated by the holomorphic sections of the normal bundle
$  H^0(D, N_{D})$. Even though by definition these deformations preserve the holomorphy of the embedding,
it is important to realize that part of them can also produce a deformation 
of the complex structure induced on $D$ by the bulk. This effect originates from the non-holomorphic split of the short exact sequence
\be\label{normalseq}
0\rightarrow T_D\rightarrow T_\X|_D\rightarrow N_D\rightarrow 0
\ee
Namely, not every holomorphic section of $N_D$ can be uplifted to a holomorphic section of $T_\X|_D$. Take the long exact sequence
\be
\ldots \rightarrow   H^0(D,T_\X|_D) \stackrel{m}\rightarrow   H^0(D,N_D) \stackrel{\delta}\rightarrow   H^1(D,T_D)\rightarrow \ldots
\ee
Since Im$\,m={\rm Ker}\,\delta$, one can see that the holomorphic sections of $N_D$ which cannot be uplifted to holomorphic sections of $T_\X|_D$ are the ones whose $\delta$-image in  $  H^1(D,T_D)$ does not vanish.

In appendix \ref{app:extrcurv} it is shown how the extrinsic curvature provides an explicit realization of the $\delta$-map. Namely, working in terms of the associated smooth complex bundles,  given a covariantly holomorphic section $V$ of $N_D$ (regarded as a U(1) line bundle), one finds that a representative of  $ H^1(D,T_D)\simeq H^{0,1}_{ \delbar }(D,T_D)$ is provided by $\Upsilon_mV^m$, where $\Upsilon_m$ is defined in (\ref{Upsilon}).

Before proceeding, notice that one can analogously consider the twisted short exact sequence
\be
0\rightarrow T_D\otimes \call_Q|_D\rightarrow (T_\X\otimes \call_Q)|_D\rightarrow N_D\otimes \call_Q|_D\rightarrow 0
\ee
and repeat all the arguments above. Take a covariantly holomorphic section  $V$  of $N_D\otimes L_Q$, which corresponds to a holomorphic section of $N_D\otimes L_Q$. Then the obstruction for the latter to become a holomorphic section of $ (T_\X\otimes \call_Q)|_D$ is given by a non-vanishing cohomology class of $\Upsilon_m V^m$ in  $ H^{0,1}_{ \delbar }(D, T_D\otimes L_Q)\simeq   H^1(D,T_D\otimes \call_Q)$.

The above arguments are clearly valid for the complex-conjugated anti-holomorphic quantities or if one replaces $\call_Q$ and $L_Q$ by any other line bundles on $\X$.

Take now the world-volume fermions $\rho$ and $\tilde\rho$. We have seen that, for them to be zero-modes, they must define harmonic representatives of $H_{\del }^{2,0}(D)$ and $H_{\delbar }^{0,2}(D,L^{-1}_Q)$ respectively. By using $\Omega$, one can associate with them the vectors $V\in H_{\delbar }^{0,0}(D,N_D\otimes L^{-1}_Q)$ and $\tilde V\in H_{\del }^{0,0}(D,\bar N_D)$ given by
\be
V^m=\frac12 e^{-\phi/2}\bar\Omega^{mab}\rho_{ab}\quad,\quad \tilde V^m=\frac12 e^{-\phi/2}\Omega^{mab}\tilde\rho_{ab}
\ee
In holomorphic indices, one then has
\be
(\Upsilon\cdot V)^{i}{}_{\bar\jmath}=-\frac12 e^{-\phi/2} \calk^{m i}{}_{\bar\jmath}\bar\Omega_{m}{}^{uv}\rho_{uv}\, ,\quad (\Upsilon\cdot \tilde V)^{\bar\imath}{}_{j}=-\frac12  e^{-\phi/2}\calk^{m\bar\imath}{}_{j}\Omega_{m}{}^{\bar u\bar v}\tilde\rho_{\bar u\bar v}
\ee
so that $\Upsilon\cdot V\in H^{0,1}_{ \delbar }(D,T_D\otimes L^{-1}_Q)$ and $\Upsilon\cdot \tilde V \in H^{1,0}_{ \del }(D, \bar T_D)$ are exactly the combinations which appear in the fermionic equations of motion (\ref{Feomflux}).

We can now turn our attention onto the world-volume flux. Supersymmetry demands it to be $(1,1)$. Clearly, any deformation of the complex structure of $D$ generated by $v\in H^{0,1}_{\delbar}(D,T_D)\simeq   H^1(D,T_D)$ can break supersymmetry since it could generate a non-vanishing
\be
\delta_v\calf^{0,2}\equiv v\cdot \calf:=v^i{}_{\bar\jmath}\calf_{i\bar u}\d \bar s^{\bar\jmath}\wedge \d s^{\bar u}
\ee
More precisely, supersymmetry is actually broken only if $\delta_v\calf^{0,2}$ cannot be reabsorbed by deforming the world-volume gauge-fields, that is only if $\delta_v\calf^{0,2}$ is  cohomologically non-trivial.

In the fermionic equations of motion, one can write
\bea
\cals_\calf\cdot \rho&\equiv& e^{-\phi/2}(\Upsilon\cdot V)\cdot \calf\equiv e^{-\phi/2}\delta_V\calf^{0,2}\cr
\tilde\cals_\calf\cdot \tilde\rho&\equiv& e^{-\phi/2}(\Upsilon\cdot \tilde V)\cdot \calf\equiv e^{-\phi/2}\delta_{\tilde V}\calf^{2,0}
\eea
Hence $\cals_\calf\cdot\rho$ and $\tilde\cals_\calf\cdot\tilde\rho$ exactly account for the generation of $(0,2)$ and $(2,0)$ components produced by the (twisted) complex structure deformations associated to  $\Upsilon\cdot V$ and $\Upsilon\cdot \tilde V$ respectively.
From this point of view, the moduli lifting conditions (\ref{fluxcohocond2}) are exactly demanding that the good zero modes should not be associated to (twisted) geometrical deformations which generate cohomologically non-trivial components $\delta_V\calf^{0,2}$ or $\delta_{\tilde V}\calf^{2,0}$, \ie\  deformations which break supersymmetry.

The question is now: are there general circumstances  under which zero-mode lifting is for sure impossible?
First of all, this happens if $\Upsilon\cdot V$ and $\Upsilon\cdot \tilde V$ are all trivial in cohomology, that is, the short exact sequence (\ref{normalseq}) splits holomorphically. Indeed,  in this case one can write
\be
\Upsilon\cdot V=\delbar_Q U \, ,\quad \Upsilon\cdot \tilde V=\del \tilde U
\ee
with $U\in T_D\otimes L^{-1}_Q$ and $\tilde U\in \bar T_D$. Then, {\em assuming} (\ref{covholBI}), one can deduce that
\be
\cals_\calf\cdot \rho=\delbar(e^{-\phi/2} U\cdot \calf) \, ,\quad \tilde\cals_\calf\cdot \tilde\rho=\del_Q(e^{-\phi/2} \tilde U\cdot \calf)
\ee
and one can see that the conditions (\ref{fluxcohocond2}) are always satisfied. If on the other hand (\ref{covholBI}) is not assumed, then one cannot reach any definitive conclusion about $\tilde\rho$.

The other special possibility is that the cohomological non-triviality of  $\calf$ is inherited from the bulk.
Remember that $e^{-\phi/2}\calf$ defines a cohomology class in $H^{1,1}_{ \delbar}(D,L_Q)$ and, {\em assuming} (\ref{covholBI}), also in  $H^{1,1}_{ \del}(D,L_Q)$. On the other hand, through the associated pull-back, the embedding $\iota:D\rightarrow \X $ allows to see part of these cohomology groups as inherited from the bulk.
In other words, ${\rm Im}\,\iota^*$ contains the classes which can be seen as the pull-back of a non-trivial class in $\X$.

Now, if $e^{-\phi/2}\calf\in {\rm Im}\,\iota^*$ in $\del_Q$ and $\delbar_Q$ cohomology, then it cannot generate any zero-mode lifting. Indeed, one can always write $\Upsilon\cdot V=\delbar_Q {\cal U}$ and  $\Upsilon\cdot \tilde V=\del \tilde {\cal U}$, where $\calu$ and $\tilde{\cal U}$ are smooth sections of the (anti-)holomorphic bundles $T_\X|_D\otimes L^{-1}_Q$ and $\bar T_\X|_D$ respectively.  Hence, if $\calf$ is the pull-back of a bulk form $\hat\calf$, one can write
\be
\cals_\calf\cdot \rho=\delbar [\iota^*(e^{-\phi/2}{\cal U}\cdot\hat\calf)] \, ,\quad
\tilde\cals_\calf\cdot \tilde\rho=\del_Q [\iota^*(e^{-\phi/2}\tilde{\cal U}\cdot\hat\calf)]
\ee
which implies that the $\rho$ and $\tilde\rho$ zero modes cannot be lifted by the flux. Again, we stress that we need (\ref{covholBI}) to arrive to such a conceivable conclusion on $\tilde\rho$.

Finally, notice that the subtleties related to (\ref{covholBI}) disappear if one only considers F-theory vacua admitting a  weak coupling orientifold limit and works on the double cover Calabi-Yau three-fold $\tilde \X$. In this case $\calf$ defines an odd cohomology class in $H^{1,1}_-(\tilde D)$, where $\tilde D$ is the double cover of $D$. The above arguments run smoothly in this case and then one can conclude that there is no flux-induced lifting of the $\rho^\alpha_{\rm z.m.}$ and $\tilde\rho^{\dot\alpha}_{\rm z.m.}$ zero modes when either the short exact sequence (\ref{normalseq}) splits holomorphically  or $\calf$ can be written as the pull-back of a two-form in $H^{1,1}_-(\tilde \X)$.

\section{Lifting zero-modes in a one-modulus example}
\label{sec:fluxexample}

In this section we will use the model developed in sections \ref{sec:example} and \ref{sec:example2}. The setting is simply $\X = \mathbb{P}^3$, and the Calabi-Yau double-cover three-fold $\tilde X$  is the octic hypersurface $\mathbb{P}^4_{1 1 1 1 4}[8]$ defined by (\ref{simpleCY}).

This provides a very simple example that shows that even in a one-modulus case, non-perturbative superpotentials can be generated.
We choose as divisor wrapped by the E3-brane the hyperplane divisor, $D\simeq H$. This corresponds  to the choice $n=1$ in the discussion of section \ref{sec:example2}. Hence, from the general results (\ref{RHspectrum}) and (\ref{LHspectrum}), we see that in absence of world-volume fluxes the fermionic spectrum for this specific case is
\be\label{ex-spectrum}
\begin{array}{c|c} \text{\lhdot\ fermions} &\text{\#\ zero m.\ ($\calf=0$)}\\ \hline
\lambda^\alpha&  2\times 1\\
\psi^\alpha & 0\\
\rho^\alpha & 0
\end{array}\quad \quad \begin{array}{c|c} \text{\rhdot\ fermions} &\text{\#\ zero m.\ ($\calf=0$)}\\ \hline
\tilde\lambda^{\dot\alpha}&  0\\
\tilde\psi^{\dot\alpha} & 0\\
\tilde\rho^{\dot\alpha} & 2\times 3\end{array}
\ee
Hence, in addition to the two universal zero modes $\lambda_{\rm z.m.}^\alpha\sim \theta^\alpha$, there are $2\times 3$ (the factor $2$ counts the two Weyl indices) zero modes $\tilde\rho_{\rm z.m.}^{\dot\alpha}$. In absence of fluxes, one has $\chi_{E3}= -2$ and then, a priori, such an instanton cannot contribute to a superpotential.

We will now show how the addition of  world-volume fluxes on $D$ can lead to the removal of the six zero modes $\tilde\rho_{\rm z.m.}^{\dot\alpha}$, so that   the divisor can contribute  to the superpotential. Actually,  as it should be clear from the general discussion of section \ref{sec:fluxduality}, describing a general world-volume flux $\calf$ is difficult, as it can undergo non-trivial monodromies, and hence cannot be treated as curvatures of ordinary line bundles or as ordinary two-forms. This difficulty is analogous to the difficulty in describing worldvolume self-dual 3-form fluxes on the dual M5-instanton explicitly in a concrete situation.

On the other hand, via the weak coupling orientifold limit, these difficulties are greatly alleviated, basically because they can be addressed in a perturbative IIB string theory picture. In that case, the purely worldvolume fluxes $F_{\rm E3}$ on the double-cover divisor $\tilde D$ are closed, orientifold-odd two-forms of $(1,1)$-type, i.e. elements of $H^{1,1}_-(\tilde D)$. Hence, via Poincar\'e duality they can be understood as divisor classes on $\tilde D$, i.e. holomorphic curves:\footnote{There is a subtlety, related to the Freed-Witten  quantization condition, that reads $\calf+\iota^*B_{\it 2}+\frac12 c_1(K_{\tilde D})\in H^2(\tilde D,\mathbb{Z})$ \cite{Minasian:1997mm, Freed:1999vc}. In our case $c_1(K_{\tilde D})=\iota^*[H]$. Since $[H]\in H^2_+(\tilde\X;\mathbb{Z})$,  one must turn on an {\em even} $[B^+_{\it 2}]=\frac12 [H]$ to cancel it, which is consistent with the orientifold action because of the periodic identification $[B^+_{\it 2}]\sim [B^+_{\it 2}]+[\omega_{\it 2}]$ for any $[\omega_{\it 2}]\in H^2(\tilde \X,\mathbb{Z})$, and then in particular for $[\omega_{\it 2}]=-[H]$.}
\be\label{fluxlift}
\frac{F_{\rm E3}}{2\pi} = \tfrac{1}{2}[\tilde D] + \sum_i n_i \big([C_i]-[C_i'] \big)\,, \quad n_i \in \mathbb{Z}
\ee
where the $[C_i]$ and $[C_i']$ are the Poincar\'e duals of holomorphic curves $C_i\subset \tilde D$ and their orientifold images, respectively. We suppress pullback symbols for simplicity. In order for the flux to survive the orientifold projection $\sigma^*(\calf_{E3}) = -\calf_{E3}$, where $\sigma$ is the involution, we turn on a B-field $B_{\it 2} = \tfrac{1}{2}\,[H]$, such that
\be
\calf =  (2\pi \ell_s)^2\sum_i n_i \big([C_i]-[C_i'] \big)\,, \quad n_i \in \mathbb{Z}
\ee

Furthermore, the primitivity condition $\calf\wedge J=0$ translates into the integrated condition
\be\label{primcond}
 \sum_i n_i \int_{C_i}J =0
\ee

Such curves can be explicitly constructed and hence provide with a very hands-on way of introducing fluxes. This approach was pioneered in the context of black hole microstate counting in \cite{Gaiotto:2005rp,
Gaiotto:2006aj,Gaiotto:2006wm}, and further developed and exploited in \cite{Denef:2007vg, Collinucci:2008pf, Collinucci:2008ht}.

The key point is that, as explained in section \ref{sec:lift}, the flux-induced lifting of the zero modes $\tilde \rho^{\dot\alpha}_{\rm z.m.}$ have a clear geometric counterpart in terms of lifting of geometric moduli. In particular, only world-volume fluxes that cannot be seen as the pull-back of some $(1,1)$ flux on the bulk can contribute. Hence, only the purely world-volume flux $F_{\rm E3}$ really matters.
Then, instead of directly calculating the action of the operator $\tilde\cals_\calf$, one can Poincar\'e-dualize the problem and look for world-volume fluxes (\ref{fluxlift}) that `rigidify' the divisor $D$. Geometrically, this will boil down to requiring that the holomorphic curves $C_i$ cannot be deformed together with $D$ while preserving their holomorphy .

Let us first discuss the geometric deformations more in detail. In the orientifold limit, they are associated to
the $H^2_-(\tilde D)$ cohomology. They correspond via Serre duality to  $H^0_+(H, \calo_{\tilde \X}(1))$, \ie\ to the sections of the normal bundle $N_{\tilde D}\simeq \calo_{\tilde \X}(1)|_{\tilde D}$ that are \emph{even} under the orientifold involution. These can be seen quite directly by writing down the equation for a generic hyperplane divisor $H$ in $\tilde \X=\mathbb{P}^4_{1 1 1 1 4}[8]$. Let the coordinates of the three-fold  be as defined earlier, $[z_1: \ldots : z_4: \xi]$, where $\xi \rightarrow -\xi$ under the involution, and $\xi$ has projective weight four. Then the E3 divisor $\tilde D$ is given by:
\be \label{generichyperplane}
P_{D} = a_1\,z_1 + \ldots + a_4\,z_4 = 0
\ee
The vanishing locus of this equation is invariant under rescalings, hence the moduli space of this divisor is a $\mathbb{P}^3$. Hence, $\text{dim }H_+^0(H, \calo_{\tilde \X}(1)) = 3$, that indeed gives the $2\times 3$ zero modes $\tilde\rho^{\dot\alpha}$. Notice however that, because $\xi$ cannot appear at degree one, all deformations will keep the divisor invariant, \ie\ as an O$(1)$ instanton. This explains why  $H^2_+(\tilde D,\calo_{\tilde D})$ is trivial and there are no zero modes $\rho^\alpha_{\rm z.m.}$.

In order to create fluxes that can rigidify the divisor, one needs to identify holomorphically embedded curves in $\tilde \X$  that are rigid, \ie\ that do not admit holomorphic deformations. The reason is the following. If one identifies fluxes on $\tilde D$ with their Poincar\'e dual by imposing that the divisor \emph{contain} assigned curves, then the condition $\calf^{0,2}=0$ translates into requiring that the curves remain holomorphic when the divisor moves. If the curves are rigid, then some or all of the divisor moduli will become obstructed.

Take the double-cover $\tilde \X$ defining equation (\ref{simpleCY}) in $\mathbb{P}^4_{1 1 1 1 4}$ to be given by:
\be\label{CYhyper}
z_1^8+ \ldots + z_4^8+\xi^2+ \psi^2\,\big(P_{4}(z) \big)^2 = 0\,,
\ee
for $\psi \in \mathbb{C}$, and $P_4(z)$ some generic polynomial of homogeneous degree four in the $z_I$. At a generic locus in complex structure moduli space, this Calabi-Yau is known to contain $29504$ isolated holomorphic degree one curves of genus zero, i.e. $\mathbb{P}^1$'s \cite{Font:1992uk, Klemm:1992tx}. We turned on the $\psi^2$ deformation in order to ensure genericity.

The curves we are interested in cannot be written as the complete intersection of the Calabi-Yau hypersurface (\ref{CYhyper}) and other two equations, regardless of whether one of these is the E3-divisor hypersurface (\ref{generichyperplane}). Indeed, such curves would be necessarily not rigid, as they would  be holomorphic regardless of the position of the E3-brane. Instead, they must be defined as complete intersections of three equations in the ambient weighted projective space. Take the following rational (genus zero) degree one curve:
\be
C_1: \quad z_1 = \eta\,z_2 \quad \cap \quad z_3 = \tilde \eta\, z_4 \quad \cap \quad \xi = \psi\,P_4 \qquad \subset \quad \mathbb{P}^4_{1 1 1 1 4}\,,
\ee
where $\eta^8 = \tilde \eta^8=-1$ are eighth roots of minus one. This curve is a $\mathbb{P}^1$, and it clearly lies inside $\tilde \X$ since it automatically satisfies  (\ref{CYhyper}). It can be shown via exact sequences that its normal bundle in $\tilde \X$  has no sections, \ie\ that
\be
N_{C_1/\tilde \X} = \calo_{\mathbb{P}^1}(-1) \oplus \calo_{\mathbb{P}^1}(-1)
\ee
In other words, this curve is fully obstructed already at first order. Its orientifold image is simply given by
\be
C_1': \quad z_1 = \eta\,z_2 \quad \cap \quad z_3 = \tilde \eta\, z_4 \quad \cap \quad \xi = -\psi\,P_4\qquad \subset \quad \mathbb{P}^4_{1 1 1 1 4}\,,
\ee
and is clearly also obstructed at first order. If one now defines the flux on the E3 as follows:
\be
\frac{F_{\rm E3}}{2\pi} = \tfrac{1}{2}\,[H] + [C_1] - [C_1']
\ee
then $ (2\pi \ell_s)^{-2}\calf \in H^{1,1}_-\cap H^2(H, \mathbb{Z})$. Requiring that this flux remain of type $(1,1)$ as the E3 deforms amounts to requiring that these two rigid curves be contained in the divisor. This imposes the following linear constraints on the $a_i$ coefficients of the divisor \eqref{generichyperplane}:
\be
a_1\,\eta+a_2=0 \quad \cap \quad a_3\,\tilde\eta+ a_4=0
\ee
The hyperplane now looks like:
\be
P_D (C_1)= a_1\,(z_1-\eta\,z_2) + a_3\,(z_3-\tilde \eta\,z_4) = 0
\ee
where by $P_D(C_1)$ we mean the most generic polynomial for a divisor $D$ subject to the constraint of containing $C_1$.
Hence, of the three moduli, two have been lifted: $h^{0,2}_- = 3 \mapsto 1$.

Since the $\tilde \X$  is a one-modulus CY, $C_1$ and $C_1'$ are homologous in $\tilde \X$, hence their difference is trivial in $H_2(\tilde \X, \mathbb{Z})$ and then in particular (\ref{primcond}) is automatically satisfied. However, the difference $C_1-C_1'$ is not trivial on the divisor. To check this, it is sufficient to find a single non-zero intersection number of this difference with some curve on the E3. Let us compute the following intersection number
\be
I = \left([C_1]-[C_1'] \right) \cdot [C_1]
\ee

The first term can be computed via the adjunction formula for the tangent bundle of the curve in terms of $[\tilde D]$ and the Poincar\'e dual $[C_1]$:
\be
\chi(C_1) = -\int_{\tilde D} \left([\tilde D] \wedge [C_1]+[C_1]\wedge [C_1] \right)
\ee
The first term is simply the intersection number $\tilde D \cdot C_1=H \cdot C_1=1$  since the curve has degree one. The fact that it is a $\mathbb{P}^1$ tells us that $\chi=2$, hence one deduces that
\be
C_1\cdot C_1 = -3
\ee
In order to get the intersection number $C_1\cdot C_1'$, one first notices that both curves have two linear equations in common. Then, one imposes two more equations of degree $4$ to intersect the curves. Hence, on this weighted projective space, one needs
\begin{eqnarray}
C_1 \cdot C_1' &=& \int_{\mathbb{P}^4_{1 1 1 1 4}} [H] \wedge [H] \wedge [4\,H]\wedge [4\,H] =4
\end{eqnarray}
Therefore
\be
I = -7
\ee
Hence, we conclude that
\begin{eqnarray}
i_*(C_1-C_1') &=& 0 \in H_2(\tilde X, \mathbb{Z}) \quad \text{but} \cr
C_1-C_1' & \neq & 0 \in H_2(\tilde D, \mathbb{Z})
\end{eqnarray}

In Poincar\'e dual terms, this implies that the flux cannot be seen as the pull-back of a (closed) flux on $\tilde \X$. This is exactly what is required for the zero-mode lifting mechanism to work, as discussed in general in \ref{sec:lift}.

Since, the class $[C_1]-[C_1']$ is non-trivial in the divisor, these two curves will not somehow recombine and annihilate.

\vskip 3mm
Now that two zero-modes have been eliminated, the remaining one can be lifted by adding another rigid curve to the [Poincar\'e dual to the] flux. Take the following curve and its orientifold image:
\be
C_2, C_2': \quad z_1 = \eta\,z_4 \quad \cap \quad z_3 = \tilde \eta\, z_2 \quad \cap \quad \xi = \pm\psi\,P_4 \qquad \subset \quad \mathbb{P}^4_{1 1 1 1 4}
\ee
Requiring that the E3-brane contain these curves will impose further restrictions on its divisor moduli that read
\be
a_1\,\eta +a_4 = 0 \quad \cap \quad a_3\,\tilde \eta+a_2=0
\ee
The divisor under consideration is described by
\be
P_D (C_1,C_2)=a( z_1-\eta\,z_2+\tfrac{\eta}{\tilde \eta}\,z_3-\eta\,z_4)  = 0
\ee

In conclusion, the flux
\be
\frac{F_{\rm E3}}{2\pi} =\tfrac{1}{2}\,[H]+ [C_1]+[C_2]-[C_1']-[C_2']
\ee
fully freezes the E3-instanton, thereby lifting all of its $2\times 3$ $\tilde\rho_{\rm z.m.}^{\dot\alpha}$ zero-modes.\footnote{One may wonder, whether the intersecting curves $C_1$ and $C_2$ might recombine into a degree two $\mathbb{P}^1$. However, for a generic hypersurface equation, such degree two curves are known to come in finite discrete amounts. In this case there are $128834912$ of them \cite{Huang:2006hq}, meaning they must be rigid. }
One can again check that the resulting linear combination of curves is non-trivial in the homology of the divisor.

Let us address one subtlety that is fortunately automatically absent in one-modulus cases. E3-instantons with fluxes could in principle have the so-called chiral zero-modes localized at intersections with D7-branes, as pointed out in \cite{ralphchiral}. However, the index that counts the next chirality of such zero-modes
\be
\int_{D7\, \cap \, E3} \left(F_{E3}-F_{D7}\right)
\ee
gets no additional contribution from the fluxes we have seen here. The fact that the differences of curves considered are trivial in the $\tilde X$ implies that the Poincar\'e-dual flux integrates to zero along any curve made by intersecting two divisors in $\tilde X$. Hence, the `magnetic' fluxes under consideration do not alter the chiral zero-mode problem. They simply lift neutral zero-modes.

\vskip 2mm
Finally, note that, although we used the specific form (\ref{CYhyper}) of the Calabi-Yau double-cover $\tilde \X$ to make life easier, we could add a host of terms to make it more generic, such as $(z_1-\eta\,z_2)\,(z_1-\eta\,z_4)\,P_6$. It is known that Calabi-Yau threefolds generically contain finite but large amounts of degree one $\mathbb{P}^1$'s (in this case  $290540$). Hence, this zero-mode lifting effect is a generic feature, and it means that this simple one-modulus Calabi-Yau admits E3-induced non-perturbative superpotentials. Since Calabi-Yau three-folds typically contain rigid curves, this phenomenon can be applied to many more situations. This opens up a new possibility for model-building. One no longer needs to construct Calabi-Yau four-fold containing exceptional divisors to put their M5-instantons on. Instead, one can study simple spaces, and discover that they already contain divisors that get frozen by such easily described fluxes.



\vspace{2cm}

\centerline{\large\em Acknowledgments}

\vspace{0.5cm}

\noindent We would like to thank L.~Anderson, I.~Brunner, V.~Braun, F.~Fucito, J.~Gray, S.~Kachru, F.~Morales, R.~Richter, T.~Weigand for useful discussions. L.~M.\ would like to thank P.~Koerber for collaboration on a related project. The work of M.~B.\ and L.~M.\ was partially supported by the ERC Advanced Grant n.226455 "Superfields", by the Italian MIUR-PRIN contract 20075ATT78,
by the NATO grant PST.CLG.978785. The work of A.~C.\ is supported in part by the Cluster of Excellence ``Origin and Structure of  the Universe'' in M\"unchen, Germany, and by a EURYI award of the European Science Foundation. A.C. would like to thank the Kavli Institute for Theoretical Physics for its hospitality during early stages of this project. L.~M.\ would like to thank  the
Laboratoire de Physique Th\'eorique et Hautes Energies for its hospitality during the course of this work.

\vspace{3cm}

\newpage

\centerline{\LARGE \bf Appendix}
\vspace{0.5cm}

\begin{appendix}

\section{U(1)-bundles and holomorphic line bundles}
\label{sec:bundles}

A U$(1)$-connection with (1,1) field-strength associated to a smooth complex line bundle $L$ allows to reinterpret $L$ as a holomorphic line bundle, that we denote as $\call$, and viceversa. Let us explicitly apply this well known result to the present case. The following discussion is completely standard.

Let us define a complex line bundle $L^q_{Q}$ with charge $q$ under the $U(1)_Q$ symmetry. Clearly, $L_Q\equiv L^{q=1}_Q$ and $L^q_{Q}=(L_Q)^q$. Hence, if $f$ is a section of $L^q_Q$, going from one patch to the other  it undergoes a transformation
\be
f\rightarrow e^{\ii q\arg(c\tau+d)}f
\ee
Its covariant derivative is given by
\be
\nabla_Q f:=(\d-\ii q Q)f=: \del_Q f+\delbar_Q f
\ee

Now, since $(\d Q)^{0,2}=0$, $Q$ defines a holomorphic line bundle.  Indeed, define
\be
\hat f= (\Im\tau)^{-q/2} f
\ee
Since, going from one patch to the other we have
\be
\Im\tau\rightarrow |c\tau+d|^{-2}\Im\tau
\ee
it is easy to see that $\hat f$ has the following gluing conditions
\be
\hat f \rightarrow (c\tau+d)^q \hat f
\ee
Hence, we see the transition functions of $\hat f$ are holomorphic and then define an associated holomorphic line bundle, that we denote
by $\call_Q^q\equiv (\call_Q)^q$. Clearly, the covariant anti-holomorphic derivative $\delbar_Q f$ is mapped into the anti-holomorphic derivative $\delbar \hat f$ and then
\be
\delbar_Q f=0 \quad\Leftrightarrow \quad \delbar\hat f=0
\ee

For instance, using this terminology, we immediately recognize that $\Omega$ as defined in (\ref{omega}) can be seen as a holomorphic section of $\call_Q\otimes K_{\X}$, whereas $e^{-\phi/2}\Omega$ is a covariantly constant section of $L_Q\otimes K_{\X}$.

Finally, notice that from $L^q_Q$ we can alternatively construct an anti-holomorphic line bundle $\bar\call^{-q}_Q$, whose sections  can be defined as $\tilde f=(\Im \tau)^{q/2} f$. The sections $\tilde f$ of $\bar\call^q_Q$  have transition functions of the form
\be
\tilde f \rightarrow (c\bar \tau+d)^{q} \tilde f
\ee
In this case $\del_Q f\rightarrow \del \tilde f$ and covariantly anti-holomorphic sections of $L^q_Q$ are mapped into anti-holomorphic sections of $\bar\call^{-q}_Q$.

Clearly, we have the isomorphisms  $L^{-q}_Q\simeq (L^q_Q)^{-1}\simeq (L^{q}_Q)^*$, $\call^{-q}_Q\simeq (\call_Q^{q})^{-1}$ and  $\bar\call^q_Q\simeq (\call_Q^{q})^*$.


\section{Some useful properties of the extrinsic curvature}
\label{app:extrcurv}

Here we review some useful properties of the extrinsic curvature.

Consider a space $M$ with metric $g$ and a submanifold $D$ with induced metric $h=\iota^*g$. We can accordingly split $T_M|_D=T_D\oplus T^\perp_D$. Then, the extrinsic curvature $\calk$ can be seen as a map
\be
\calk: T_D\otimes T_D\rightarrow T^\perp_D
\ee
defined by
\be\label{ec1}
\calk(v,w)=\calk(w,v)=(\hat\nabla_v w)^\perp\qquad \forall v,w\in T_D
\ee
where $\hat \nabla$ is computed by using the bulk metric $g$.
Alternatively
\be\label{ec2}
\langle X,\calk\rangle\equiv X_m\calk^m=-\frac12(\call_X g)|_D\qquad \forall X\in T^\perp_D
\ee
that can be easily proved by starting from the definition (\ref{ec1}).

Assume now that $M$ is K\"ahler, with complex structure $I$ and K\"ahler form $J=g\cdot I$.  We have in particular that $I$ and $J$ are covariantly constant with respect to the metric $g$.  Furthermore, let us restrict to holomorphic submanifolds $D$.
This latter condition can be written by saying that
\be
I\cdot v\subset T_D\qquad\forall v\in T_D
\ee
Then,
\be\label{holindex}
\calk(Iv,w)=\calk(v,Iw)=I\calk(v,w)
\ee
which means that $\calk$ has pure holomophic or anti-holomorphic indices.

Notice also that, since $\calk$ has pure indices, the definition (\ref{ec2}) is telling us that $\calk$ measures the deformation of the complex structure on $D$ induced by a deformation of $D$ preserving the holomorphy of the embedding. Namely, suppose that $V$ is a section of the ordinary normal bundle that generates  such a deformation.

Then, the variation of the induced metric generated by $V$ is given by
\be
\delta_V h=(\call_V g)|_D=-2V_m\calk^m
\ee
so that
\be
(\delta_V h)_{i\bar\jmath}=0\quad,\quad (\delta_V h)_{\bar\imath\bar\jmath}=-2V_{m}\calk^{m}{}_{\bar\imath\bar\jmath}
\ee
Notice that the geometric deformation does not induce any deformation of the K\"ahler structure on $D$. One can alternatively parametrize the change in the complex structure on $D$ by a section of $(\delta_V I_D)^{i}{}_{\bar\jmath}\in\Gamma(T_D\otimes \bar T^*_D)$, where $T_D$ and is the holomorphic tangent bundle. $I_D+\delta_V I_D$ denotes the new complex structure\footnote{The condition $(I_D+\delta_V I_D)^2=-\bbone $ implies that $\delta I_D$ has no pure indices.} in which $h+\delta h$ is hermitian. Hence
\be\label{ik}
 h_{ac}(\delta_V I_D)^c{}_b+ h_{bc}(\delta_V I_D)^c{}_a=\ii \delta h_{ab}=-2iV_m\calk^m{}_{ab}
\ee
Since $X$ preserves the holomorphy of the embedding, in complex coordinates $(s^i,\bar s^{\bar\imath})$ we have
\be\label{CSintcond}
\delbar_{[\bar\imath}(\delta_VI_D)^{j}{}_{\bar u]}=0
\ee
and hence we can locally write
\be
(\delta_VI_D)^{i}{}_{\bar\jmath}=\delbar_{\bar\jmath}\,\delta_V\zeta^{i}
\ee
where $\delta_V\zeta^{i}$ defines the shift in the new complex coordinates $\tilde\zeta^{i}=\zeta^i+\delta_V\zeta^{i}$.
Using this one can show that $ h_{ac}(\delta_V I_D)^c{}_b$ is symmetric in $(ab)$. Hence from (\ref{ik}) we get
\be\label{defK}
(\delta_V I_D)^i{}_{\bar\jmath}=-\ii h^{i\bar u}V_m\calk^m{}_{\bar u\bar\jmath}
\ee

We can use the extrinsic curvature to define the operator
\be\label{Upsilon}
{\Upsilon}: T^\perp_D\rightarrow T^{1,0}_D\otimes T^{*0,1}_D\qquad\qquad \text{with}\quad (\Upsilon_m)^i{}_{\bar\jmath}:= -\ii h^{i\bar u}g_{mn}\calk^n{}_{\bar u\bar\jmath}
\ee
In particular, the action of  $\Upsilon$ descends to an action at the level of sheaf cohomology classes
\be\label{cohoUpsilon}
\Upsilon: H^{0}(D,N_D)\rightarrow H^{1}(D,T_D)\
\ee
where $N_D$ is the holomorphic normal bundle to $D$. In the form (\ref{cohoUpsilon}), $\Upsilon$ gives precisely the map that associates  any infinitesimal deformation of the divisor with the corresponding infinitesimal deformation of complex structure on $D$.


\end{appendix}




\end{document}